\documentclass[10pt]{article}
\usepackage[margin=0.8in]{geometry}
\usepackage{authblk}
\usepackage{sectsty}
\usepackage{amsmath}
\usepackage{graphicx}
\usepackage{enumerate}
\usepackage{natbib}
\usepackage{url}
\usepackage{hyperref}
\usepackage{amsfonts,amssymb}
\usepackage{subfig,etoolbox}
\sectionfont{\large}
\subsectionfont{\large}
\numberwithin{equation}{section}
\newtheorem{def1}{Definition}
\numberwithin{def1}{section}
\newtheorem{prop1}{Proposition}
\numberwithin{prop1}{section}
\newtheorem{th1}{Theorem}
\numberwithin{th1}{section}
\newtheorem{res1}{Result}
\numberwithin{res1}{section}
\newtheorem{alg1}{Algorithm}
\numberwithin{alg1}{section}
\newcommand\blfootnote[1]{%
  \begingroup
  \renewcommand\thefootnote{}\footnote{#1}%
  \addtocounter{footnote}{-1}%
  \endgroup}
\title{\fontsize{15}{18}\selectfont \textbf{Modeling temporal dependence of longitudinal data: use of multivariate geometric skew-normal copula}}
\author[]{\fontsize{10}{12}\selectfont Subhajit Chattopadhyay}
\affil[]{\small Bhubandanga, Bidya Sagar Path, Bolpur, West Bengal 731204, India}
\date{}
\begin{document}
\maketitle
\begin{abstract}
The utilization of copulas for modeling dependence has garnered significant attention in recent years. Conversely, the quest for multivariate copulas with desirable dependence properties remains a crucial area of research. When fitting regression models to longitudinal data, the multivariate Gaussian copula is frequently employed to accommodate the temporal dependence of repeated measurements. However, using a symmetric multivariate Gaussian copula may not be ideal in all scenarios, as it fails to capture non-exchangeable dependence or tail dependence if present in the data. Therefore, to ensure reliable inference, it is imperative to explore beyond the Gaussian dependence assumption. In this paper, we introduce the construction of a geometric skew-normal copula derived from the multivariate geometric skew-normal (MGSN) distribution proposed by \citet{kundu2014geometric} and \citet{kundu2017multivariate}, aimed at modeling the temporal dependence of non-Gaussian longitudinal data. Initially, we examine the dependence properties of the proposed multivariate copula and subsequently develop regression models for both continuous and discrete longitudinal data. Notably, the quantile function of this multivariate copula remains independent of the correlation matrix of its respective multivariate distribution, offering computational advantages in likelihood inference compared to the copulas derived from skew-elliptical distributions as proposed by Azzalini and others. Furthermore, composite likelihood inference becomes feasible for this multivariate copula, allowing for parameter estimation from ordered probit models with the same dependence structure as the geometric skew-normal distribution. We conduct extensive simulation studies to validate our proposed models and apply them to analyze the longitudinal dependence of two real-world datasets. Finally, we present our findings in terms of the improvements over regression models based on multivariate Gaussian copulas. 
%% Updated April 2024. %%
\end{abstract}
\blfootnote{\textbf{Correspondence}: Subhajit Chattopadhyay, email: subhajitc.stat.isi@gmail.com, orchid: 0000-0002-6893-9132}
\section{Introduction}
Analysis of longitudinal data using regression models has been extensively addressed in the statistical literature (see, for instance, \citet{liang1986longitudinal} or \citet{fitzmaurice2008longitudinal}). Many of these models rely on the assumption of multivariate normal (MVN) distribution. However, this assumption of normality for the repeated measurements may not always be appropriate, particularly when standard graphical diagnostics such as histograms or pairwise scatter plots indicate asymmetry in the marginal distributions as well as in the dependence pattern over time. Furthermore, these models are not suitable for analyzing discrete longitudinal data. There is a growing demand among practitioners for flexible models that can handle such data, which should be computationally tractable and possess useful dependence properties while maintaining simple marginal interpretations. Copulas offer a flexible and effective framework for modeling dependent repeated measurements of any type. In typical applications, marginal distributions for each repeated measurement are arbitrarily chosen, and the dependence between them is specified through a parametric copula function.
\par Elliptical copulas, such as multivariate Gaussian or Student-$t$, are widely favored in the realm of multivariate dependence due to their straightforward parametric inference, as highlighted by \citet{xue2000multivariate}, \citet{masarotto2012gaussian}, and \citet{sun2008heavy}. They serve as building blocks for constructing parametric models accommodating both continuous data with various marginal distributions and discrete data within latent variable frameworks. \citet{joe1997multivariate} and \citet{nelsen2006introduction} provide comprehensive discussions on copula models and their dependence properties. However, Gaussian copulas, despite their popularity, fall short in capturing dependence among non-exchangeable and extreme variables, as pointed out by \citet{ang2002asymmetric} and \citet{patton2006modelling}. \citet{demarta2005t} proposed employing multivariate Student-$t$ and related copulas to address tail dependence in correlated data. While copula-based models often lack closed-form expressions and necessitate numerical computations, most commonly used multivariate copulas assume exchangeability, rendering them unable to distinguish varying influences among variables. Alternative approaches have been taken in the literature by utilizing copulas generated from asymmetric multivariate distributions described in \citet{azzalini2013skew} to distinguish between non-exchangeable events. But these copulas have limitations in applications such as lack of algebraically tractable forms, and difficulties in terms of parametric inference. \citet{wei2016multivariate} utilized skew-normal copulas with block coordinate algorithms for parameter estimation to capture non-exchangeable dependence. \citet{smith2012modelling} discussed Bayesian inference and applications of skew-$t$ copulas, while \citet{yoshiba2018maximum} addressed the numerical challenges in obtaining maximum likelihood estimates for skew-$t$ copulas in high dimensions. Furthermore, a limitation of these multivariate copulas is that their parameters cannot be uniquely identified from their lower-dimensional densities.
\par Recently, \citet{kundu2014geometric} introduced an alternative skew-normal distribution, extended to the multivariate domain in \citet{kundu2017multivariate}, where the multivariate normal distribution serves as a special case. Termed the geometric skew-normal (GSN) distribution, this model arises as a geometric sum of independent and identically distributed normal random variables. Differing from Azzalini's skew-normal distribution, the GSN distribution can exhibit multimodality and various shapes contingent upon its three-set parameter values. The author has derived several intriguing properties of this distribution and demonstrated its computational ease in multivariate scenarios. Given its novelty, the literature offers relatively few applications of this distribution. \citet{roozegar2017power} explored a category of power series skew-normal distributions by extending the GSN distribution. \citet{redivo2020bayesian} proposed a Bayesian model-based clustering method using the geometric skew-normal distribution and validated its performance through simulation studies. In this study, we propose an alternative asymmetric multivariate copula derived from the geometric skew-normal distribution to model the temporal dependence of non-Gaussian longitudinal data. Initially, we derive the dependence properties of the proposed copula and subsequently develop appropriate dependence models for continuous and ordinal longitudinal data. For continuous repeated measurements, we employ generalized linear models for the marginals, while for ordinal responses, we adopt a latent variable formulation. Notably, the quantile function of this multivariate copula remains independent of the correlation matrix of its respective multivariate distribution, offering computational advantages in parametric inference. Another noteworthy advantage over Azzalini's skew-normal copula is that the multivariate GSN copula is closed under marginalization, implying that all its lower-dimensional sub-copulas belong to the same parametric family. Consequently, composite likelihood inference becomes feasible for this multivariate copula, facilitating parameter estimation in ordered probit models with a dependence structure governed by the geometric skew-normal distribution.
\par Rest of the article is organized as follows. In Section \ref{sec2}, the details of construction of the multivariate geometric skew-normal copula are described. Section \ref{sec3} elaborates the dependence properties of the GSN copula. The details of maximum likelihood estimation for unrestricted GSN copula using block-coordinate ascent algorithm is described in Section \ref{sec4}. In Section \ref{sec5}, we develop regression models for continuous and ordinal longitudinal data and describe their parametric inference. In Section \ref{sec6}, we describe some standard model evaluation methods. Section \ref{sec7} presents the finite sample performance of our proposed models using some simulated data sets. Thereafter in Section \ref{sec8}, we analyze two real world data sets and compare the fits with corresponding Gaussian copula based models. Section \ref{sec9} concludes this article with a general discussion.
%% Section 2. %%
\section{Geometric skew-normal copula}\label{sec2}
In this section we discuss in details the construction of multivariate geometric skew-normal copula. A $d$-variate multivariate distribution function $C(u_1,\dots,u_d)$  is called a copula if its marginal distributions are uniformly distributed on $[0,1]$. Let $\mathbf{X} = (X_1,\dots,X_d)^\intercal$ be a random vector with monotone marginal distribution functions $F_i(x_i)$, for $i = 1,\dots,d$ and the density functions $f_i(x_i)$ then by Sklar's theorem there exist a copula $C : [0,1]^d \to [0,1]$ such that
\begin{equation}\label{skcop}
F(\mathbf{x}) = C(F_1(x_1),\dots,F_d(x_d)), \quad \mathbf{x} \in \mathcal{R}^d.
\end{equation}
Moreover, the copula is unique if the joint distribution function $F : \mathcal{R}^d \to [0,1]$ is continuous (see \citet{sklar1959fonctions}). The parametric copulas are generally constructed from a continuous multivariate distribution function $F(\mathbf{x},\mathbf{\theta})$ with strictly monotonic marginal distributions $F_i(x_i,\theta)$ for $i = 1,\dots,d$ and a parameter vector $\mathbf{\theta}$. Then the copula density $c$ on $[0,1]^d$ for $u_i = F_i(x_i,\theta)$ can be obtained by differentiation as
\begin{equation}\label{defcop}
c(u_1,\dots,u_d) = \frac{\partial^d C(u_1,\dots,u_d)}{\partial u_1\dots\partial u_d}.
\end{equation}
When $F(.)$ in (\ref{skcop}) is a skew-elliptical distribution function, the resulting copula is called a skew-elliptical copula. \citet{kollo2013multivariate} discussed the construction of copulas from skew-elliptical class of distributions. From equations (\ref{skcop}) and (\ref{defcop}) one can present the multivariate density of function of $\mathbf{X}$ through the copula density by
\begin{equation}\label{copden0}
f(x_1,\dots,x_d,\mathbf{\theta}^*) = c(F_1(x_1),\dots,F_d(x_d),\mathbf{\phi})\prod_{i=1}^d f_i(x_i,\mathbf{\theta}),
\end{equation}
where $\mathbf{\theta}^* = (\mathbf{\theta}^\intercal,\mathbf{\phi}^\intercal)^\intercal$.
Reversing the above relation, the copula density $c : [0,1]^d \to \mathcal{R}$ can be expressed through the densities of $\mathbf{X}$ and $X_i$ for $i = 1,\dots,d$ as
\begin{equation}\label{copden1}
c(\mathbf{u},\mathbf{\phi}) = \frac{f(F_1^{-1}(u_1),\dots,F_d^{-1}(u_d),\mathbf{\theta}^*)}{\prod_{i=1}^d f_i(F_i^{-1}(u_i),\theta)}.
\end{equation}
The copula captures dependence structure of the data via some set of parameters of $F(.)$. Here in this article we construct a copula by using geometric skew-normal distribution as described next. That is GSN copula can be seen as an one-to-one transformation from $\mathbf{X}$ having geometric skew-normal distribution. One can see \citet{joe2014dependence} for an overview of copula based modeling.
%% New subsection. %%
\subsection{Multivariate geometric skew-normal distribution}
Multivariate geometric skew-normal variable can be expressed as a geometric random sum of Gaussian random variables. A $d$-variate MGSN distribution is defined as follows.
\begin{def1}
Suppose $N \sim GE(p)$, and $\{\mathbf{X}_i; i = 1,2 \dots\}$ are i.i.d. $N_d(\mathbf{\mu,\Sigma})$ random vectors. It is assumed that $N$ and $\mathbf{X}_i$'s are independently distributed. Then the random variable $\mathbf{X}$, where
\begin{equation}\label{rep1}
\mathbf{X} \overset{dist}{=} \sum_{i=1}^N \mathbf{X_i}
\end{equation}
is said to have a $d$-variate geometric skew-normal distribution with parameters $p,\mathbf{\mu}$ and $\mathbf{\Sigma}$. We denote this distribution by $MGSN_d(p,\mathbf{\mu,\Sigma})$.
\end{def1}
From \citet{kundu2017multivariate}, if $\mathbf{X} \sim MGSN_d(p,\mathbf{\mu,\Sigma})$, then the CDF and PDF of $\mathbf{X}$ take the following forms -
\begin{equation}\label{cdf1}
F_{d,GSN}(\mathbf{x}|\mathbf{\mu,\Sigma},p) = \sum_{k=1}^\infty p(1-p)^{k-1}\Phi_d(\mathbf{x}|k\mathbf{\mu},k\mathbf{\Sigma}) \quad \text{and}
\end{equation}
%% %% %% %%
\begin{align}\label{pdf1}
f_{d,GSN}(\mathbf{x}|\mathbf{\mu,\Sigma},p) & = \sum_{k=1}^\infty p(1-p)^{k-1}\phi_d(\mathbf{x}|k\mathbf{\mu},k\mathbf{\Sigma}) \nonumber \\
& = \sum_{k=1}^\infty \frac{p(1-p)^{k-1}}{(2\pi k)^{d/2}\sqrt{|\mathbf{\Sigma}|}} e^{-\frac{1}{2k}(\mathbf{x} - k\mathbf{\mu})^\intercal \mathbf{\Sigma}^{-1}(\mathbf{x} - k\mathbf{\mu})},
\end{align}
respectively. Here $\Phi_d(\mathbf{x}|k\mathbf{\mu},k\mathbf{\Sigma})$ and $\phi_d(\mathbf{x}|k\mathbf{\mu},k\mathbf{\Sigma})$ denote the CDF and PDF of a $d$-variate normal distribution respectively, with mean vector $k\mathbf{\mu}$ and dispersion matrix $k\mathbf{\Sigma}$. The PDF of standard $MGSN_d(p)$ distribution where $\mathbf{\mu} = 0$ and $\mathbf{\Sigma} = I$ is symmetric and unimodal, for all values of $d$ and $p$. But the PDF of $MGSN_d(p,\mathbf{\mu,\Sigma})$ can be skewed and multimodal as well depending on parameter values. Note that for $p = 1$ it reduces to $N_d(\mathbf{\mu,\Sigma})$ distribution. Similarly for the univariate case we have the CDF as
\begin{equation}\label{cdfu1}
F_1(x|\mu,\sigma,p) = \sum_{k=1}^\infty p(1-p)^{k-1}\Phi\left(\frac{x - k\mu}{\sigma\sqrt{k}}\right),
\end{equation}
where $X$ follows $GSN(\mu,\sigma,p)$ when $d = 1$. Generation of random samples from multivariate geometric skew-normal distribution is simple with two steps. Now to construct the GSN copula the following results are needed.
\begin{res1}\label{res1}
If $\mathbf{X} \sim MGSN_d(p,\mathbf{\mu,\Sigma})$ and $\mathbf{X}_1 \sim MGSN_h(p,\mathbf{\mu}_1,\mathbf{\Sigma}_{11})$ where
\begin{equation}
\mathbf{X} = \left(\begin{matrix} \mathbf{X}_1 \\ \mathbf{X}_2 \end{matrix}\right), \;\; \mathbf{\mu} = \left(\begin{matrix} \mathbf{\mu}_1 \\ \mathbf{\mu}_2 \end{matrix}\right) \; \text{and} \;\; \mathbf{\Sigma} = \left(\begin{matrix} \mathbf{\Sigma}_{11} & \mathbf{\Sigma}_{12} \\ \mathbf{\Sigma}_{21} & \mathbf{\Sigma}_{22} \end{matrix}\right) \; \text{are}
\end{equation}
the partition of the vector and parameters of dimension $(h, d-h)$ then $\mathbf{X}_2 \sim MGSN_{d-h}(p,\mathbf{\mu}_2,\mathbf{\Sigma}_{22})$.
\end{res1}
%% Another result. %%
\begin{res1}\label{res2}
If $\mathbf{X} \sim MGSN_d(p,\mathbf{\mu,\Sigma})$, then $\mathbf{Z} = \mathbf{DX} \sim MGSN_s(p,\mathbf{D\mu},\mathbf{D\Sigma D^\intercal})$, where $\mathbf{D}$ is a $s\times d$ matrix of rank $s\leq d$.
\end{res1}
These are similar to the multivariate normal distribution and provide the marginals of a MGSN distribution. The joint to marginal relationship here is much simpler than Azzalini's skew-elliptical class of distributions. To construct the copula from (\ref{res1}) and (\ref{res2}), without loss of generality we can take $\mathbf{\Sigma}$ to be a correlation matrix by putting $\mathbf{D} = diag(\sigma_{11}^{-1/2},\dots,\sigma_{dd}^{-1/2})$.
%% Next we describe the copula with dependence structure. %%
\subsection{Construction of the GSN copula}
\begin{figure}
    \centering
    \includegraphics[width=14cm]{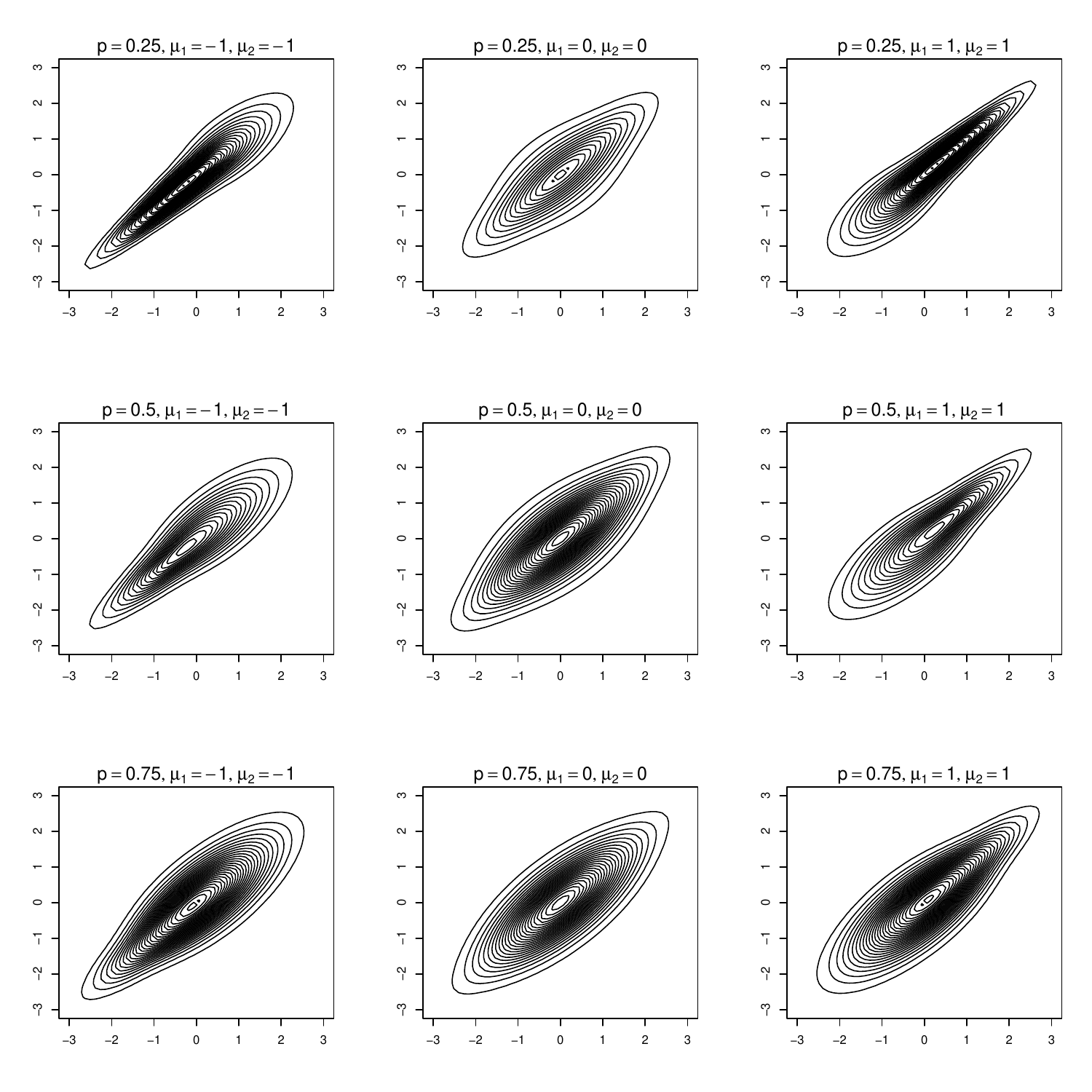}
    \caption{Contour plots of bivariate geometric skew-normal copula  using standard normal marginals. The values of the parameters are used as $p = \{0.25,0.5,0.75\}$, $\mathbf{\mu} = \{(-1,-1),(0,0),(1,1)\}$ and the common parameter $\mathbf{\rho} = 0.77$.}
    \label{fig:gsn_contour1}
\end{figure}

\begin{figure}
    \centering
    \includegraphics[width=14cm]{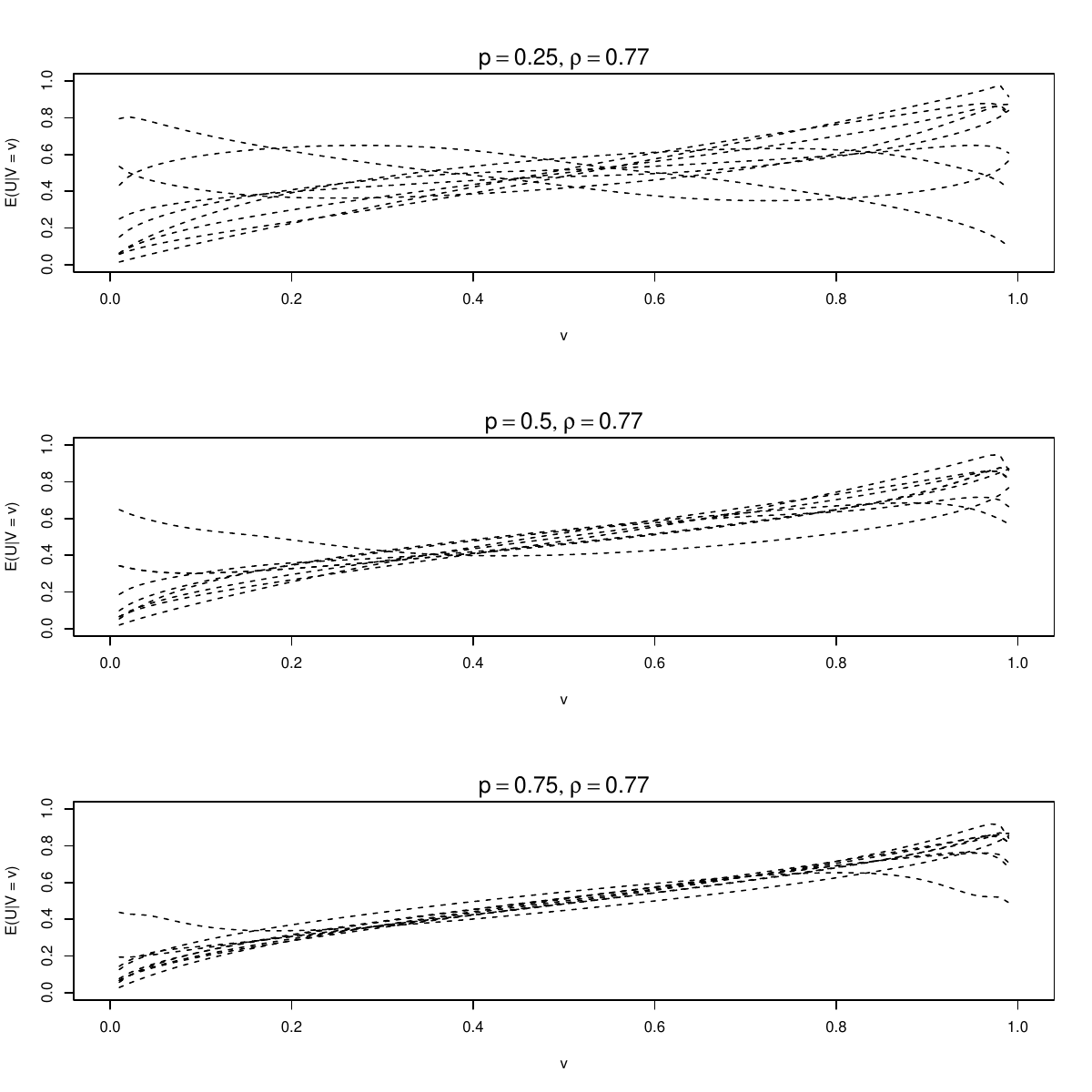}
    \caption{Regression curves of bivariate geometric skew-normal copula. The values of the parameters are used as $p = \{0.25,0.5,0.75\}, \mu = \{(-1,-1),(0,0),(1,1)\}$ and $\rho = 0.77$.}
    \label{fig:reg.gsn1}
\end{figure}

If $\mathbf{\Sigma}$ is a correlation matrix then the $j$-th marginal distribution of the $d$-variate $MGSN_d(p,\mathbf{\mu,\Sigma})$ distribution is $GSN(p,\mu_j,1)$. Now we propose the geometric skew-normal copula as follows.
\begin{def1}
A $d$-dimensional copula $C_{d,GSN}$ is called a GSN copula with parameters $\mathbf{\mu,\Sigma}$ and $p$ if
\begin{equation}\label{gsncop1}
C_{d,GSN}(\mathbf{u}|\mathbf{\mu,\Sigma},p) = F_{d,GSN}(F_1^{-1}(u_1|\mu_1,1,p),\dots,F_1^{-1}(u_d|\mu_d,1,p)|\mathbf{\mu,\Sigma},p)
\end{equation}
where $F_1^{-1}(u_j|\mu_j,1,p)$ denotes the inverse of the CDF of the $GSN(p,\mu_j,1)$ distribution and $\mathbf{\Sigma}$ denote a correlation matrix. The corresponding geometric skew-normal copula density is given by
\begin{equation}\label{copden2}
c_{d,GSN}(\mathbf{u}|\mathbf{\mu,\Sigma},p) = \frac{f_{d,GSN}(F_1^{-1}(u_1|\mu_1,1,p),\dots,F_1^{-1}(u_d|\mu_d,1,p)|\mathbf{\mu,\Sigma},p)}{\prod_{j=1}^d f_{1,GSN}(F_1^{-1}(u_j|\mu_j,1,p))}
\end{equation}
where the multivariate density $f_{d,GSN}(.)$ is given in (\ref{pdf1}) and $f_{1,GSN}$ is the marginal density of a geometric skew-normal variable $X_j \sim GSN(p,\mu_j,1)$.
\end{def1}
It is direct from (\ref{pdf1}) that Gaussian copula is nested in (\ref{gsncop1}) when $p = 1$. Thus we derive the dependence properties when $0 < p < 1$. Unlike the skew-elliptical copulas, we have no such slant parameter here and the parametric relations to the marginal distributions are simpler. To illustrate the dependence shapes imposed by the bivariate GSN copula model, in Figure \ref{fig:gsn_contour1} we provide the contour plots of the densities with $N(0,1)$ marginals. As we can see, the asymmetry in this copula comes from the location parameter $\mu$ of the bivariate normal component. Following \citet{nikoloulopoulos2009finite}, we can also graphically represent the imposed dependence of the bivariate GSN copula using regression curves based on conditional expectation. In Figure \ref{fig:reg.gsn1} we plot
\begin{equation}\label{regcurve1}
E(U|V = v) = \int_0^1 u c(u|v) du = \int_0^1 u c(u,v) du,
\end{equation}
where $U$ and $V$ are uniformly distributed random variables on the interval $(0,1)$, for various values of $p$ and $(\mu_1,\mu_2)^\intercal$ taking values in the set $\{-1,0,1\}$. The shape of the conditional expectation depends on the magnitude and the sign of the location parameter of the bivariate normal component and the mean of the geometric random variable. This shows a wide range of different relationships can be modeled by the GSN copula. It also reveals how shifting the value of $p$, from $0$ to $1$ the dependence tend to be linear which is same as the Gaussian copula. The generation of random samples from MGSN distribution is very simple and so is from the copula. A random sample from $C_{d,GSN}(p,\mathbf{\mu,\Sigma})$ can be obtained using the following algorithm.
\begin{alg1}\label{algo1}
(Sampling from the GSN copula.)
\begin{itemize}
    \item Step 1: Generate $n \sim GE(p)$.
    \item Step 2: Generate $X \sim N_d(n\mathbf{\mu},n\mathbf{\Sigma})$.
    \item Step 3: Set $U_j = F_1(X_j|\mathbf{\mu}_j,1,p)$ for $j = 1,\dots,d$.
\end{itemize}
\end{alg1}
%% The plots. %%
%% The copula contour plots. %%
%% The regression curves. %%
%% The section 3. %%
\section{Dependence properties}\label{sec3}
In this section we discuss different properties of a GSN copula. But first we need to look at the correlation structure of the MGSN distribution as discussed in \citet{kundu2017multivariate}, derived from its moment generating function. If $\mathbf{X} = (X_1,\dots,X_d)^\intercal \sim MGSN_d(p,\mathbf{\mu,\Sigma})$ where $\mathbf{\Sigma}$ is a correlation matrix then
\begin{equation}\label{corgsn1}
Corr(X_i,X_j) = \frac{p\rho_{ij} + \mu_i\mu_j(1-p) }{\sqrt{p + \mu_i^2(1-p)}\sqrt{p + \mu_j^2(1-p)}}.
\end{equation}
Hence from (\ref{corgsn1}) the correlation between $X_i$ and $X_j$ for $i\neq j$ depends on $\rho_{ij}$ as well as $\mu_i$ and $\mu_j$. It follows that when $\mu_i = \mu_j = \rho_{ij} = 0$ for fixed $p$, $Corr(X_i.X_j) = 0$. Therefore, in this case although $X_i$ and $X_j$ are uncorrelated, they are not independent. But the Pearson's correlation coefficient is a symmetric measure of dependence which doesn't tell which variable have more influence on other. Hence we need to investigate other dependence measures of the GSN copula.
\par If the dependence between two variables is such that if one variable increases then the other tends to increase or decrease, then it is referred as monotone association. The measures based on concordance and discordance such as Kendall's tau or Spearman's rho, are invariant with respect to the marginal distributions for continuous random variables, i.e. they can be expressed as a function of their copula. Multivariate extensions of such measures have been discussed by several authors as \citet{nelsen2002concordance} or \citet{joe1990multivariate}. Kendall's tau and Spearman's rho for bivariate GSN copula can be obtained as an infinite sum of bivariate normal probabilities. The population version of Kendall's tau is defined as the probability of concordance minus the probability of discordance given a bivariate random vector $(X_1,X_2)^\intercal$, i.e.
\begin{equation}
\mathbf{\tau} = P[(X_1 - X'_1)(X_2 - X'_2) > 0] - P[(X_1 - X'_1)(X_2 - X'_2) < 0], \nonumber 
\end{equation}
where $(X'_1,X'_2)^\intercal$ is independent and identically distributed as $(X_1,X_2)^\intercal$ with distribution function $F$. If $F$ has the bivariate copula $C$, this is equal to
\begin{equation}\label{kendpr1}
\mathbf{\tau}(C) = 4\int_0^1 \int_0^1 C(u,v)dC(u,v) - 1 = 4P(X'_1 < X_1,X'_2 < X_2) - 1.
\end{equation}
For a bivariate GSN copula we obtain $\mathbf{\tau}$ as follows.
\begin{prop1}\label{kend1}
Let $\mathbf{X},\mathbf{X}' \sim MGSN_2(p,\mathbf{\mu,\Sigma})$ are independent and $\mathbf{\Sigma}$ is a bivariate correlation matrix. Then Kendall's tao is given by
\begin{align}\label{kendal1}
\mathbf{\tau}(C_{GSN}) & = 4p^2\sum_{n=1}^\infty\sum_{n'=1}^\infty (1-p)^{n+n'-2}\Phi_2\Big(\frac{n'-n}{\sqrt{n+n'}}\mathbf{\Sigma}^{-1/2}\mathbf{\mu}\Big) - 1 \\
& = 4p^2\sum_{n=1}^\infty\sum_{n'=1}^\infty (1-p)^{n+n'-2} \int_{-\infty}^\frac{(n'-n)\mu_1}{\sqrt{n+n'}}\Phi\Big(\frac{(n'-n)\mu_2}{\sqrt{(n+n')(1-\rho^2)}} - \frac{\rho y}{\sqrt{1-\rho^2}}\Big)\phi(y)dy - 1. \nonumber
\end{align}
\end{prop1}
For a random vector $(X_1,X_2)^\intercal$ with continuous marginal distribution functions $F_j, j = 1,2$, Spearman's rho is defined as $\mathbf{\rho} = Corr(F_1(X_1),F_2(X_2))$. Under the copula representation,
\begin{equation}
\mathbf{\rho}(C) = 12\int_0^1 \int_0^1 C(u,v)dudv - 3 = 12P(X^*_1 < X_1,X^*_2 < X_2) - 3
\end{equation}
where $X^*_j \sim F_j, j = 1,2$, independently and aslo independent with $(X_1,X_2)^\intercal$. We can also calculate $\mathbf{\rho}$ for a bivariate GSN copula.
\begin{prop1}\label{sper1}
Let $\mathbf{X} \sim MGSN_2(p,\mathbf{\mu,\Sigma})$ and $\mathbf{X}^* \sim MGSN_2(p,\mathbf{\mu,I})$ are independent where $\mathbf{\Sigma}$ is a bivariate correlation matrix and $I$ is the identity matrix. Then Spearman's rho is given by
\begin{align}\label{spearman1}
& \mathbf{\rho}(C_{GSN}) = 12p^2\sum_{n^*=1}^\infty\sum_{n=1}^\infty (1-p)^{n^*+n-2}\Phi_2\Big((n-n^*)(n^*\mathbf{I}+n\mathbf{\Sigma})^{-1/2}\mathbf{\mu}\Big) - 3 = \\
& 12p^2\sum_{n^*=1}^\infty\sum_{n=1}^\infty (1-p)^{n^*+n-2} \int_{-\infty}^\frac{(n-n^*)\mu_1}{\sqrt{n^*+n}}\Phi\Big(\frac{(n-n^*)\mu_2\sqrt{n^*+n}}{\sqrt{(n^*+n)^2 - n^2\rho^2}} - \frac{n\rho y}{\sqrt{(n^*+n)^2 - n^2\rho^2}}\Big)\phi(y)dy - 3. \nonumber
\end{align}
\end{prop1}
It is important to point out that when $\mathbf{\mu} = (\mu_1,\mu_2) = 0$, we find $\mathbf{\tau}(C_{GSN}) = (2/\pi)\arcsin{\rho}$, which is same as the bivariate Gaussian copula. But that is not the case with $\mathbf{\rho}(C_{GSN})$ as it simplifies to
\begin{equation}
\mathbf{\rho}(C_{GSN}) = 12p^2\sum_{n^*=1}^\infty\sum_{n=1}^\infty (1-p)^{n^*+n-2} \arcsin{\frac{np}{n^*+n}}. \nonumber
\end{equation}
One can obtain the values of (\ref{kendal1}) and (\ref{spearman1}) by a numerical approximation of the infinite series up to some finite terms and a numerical integration.
\par Most of the commonly used copulas for applied research such as Archimedean and meta-elliptical copulas assume that the dependence structure between the variables of interest is symmetric. Asymmetry in the copula literature is determined by the joint upper and lower tails of multivariate distributions as mentioned in \citet{joe2014dependence}. The definitions regarding the symmetry properties of copula are given below. 
\begin{def1}\label{exch1}
A $d$-dimensional copula $C$ is exchangeable or permutation symmetric if it is the distribution function of an uniform vector $\mathbf{U} = (U_1,\dots,U_d)^\intercal$ satisfying
\begin{equation}
C(u_1,\dots,u_d) = C(u_{r(1)},\dots,u_{r(d)}) \nonumber
\end{equation}
for any permutation $r \in \Gamma$, where $\Gamma$ denotes the set of all permutations on the set $\{1,\dots,d\}$.
\end{def1}
Note that a $d$-dimensional continuous random vector $\mathbf{X}$ is exchangeable if and only if the marginal CDFs are identical and the copula is exchangeable. Most of the commonly used two-parameter bivariate copula families are exchangeable.
\begin{def1}\label{rad1}
A $d$-dimensional copula $C$ is reflection symmetric if $\mathbf{U}$ has the same distribution as $\mathbf{1 - U}$ where $\mathbf{1 - U} = (1 - U_1,\dots,1 - U_d)^\intercal$.
\end{def1}
The definition of reflection or central symmetry is that a $d$-dimensional random vector $\mathbf{X}$ as centrally symmetric about $\mathbf{a} = (a_1,\dots,a_d)^\intercal$ if, and only if each $X_i$ is marginally symmetric about $a_i$ and the corresponding copula $C$ is reflection symmetric. If $\mathbf{U} \sim C$ and $\mathbf{1 - U} \sim \hat{C}$ then $\hat{C}$ is called a reflected copula of $C$. \citet{nelsen2006introduction} called the condition of reflection symmetry, $C \equiv \hat{C}$ as radial symmetry. If the copula density as in (\ref{defcop}) exists, then reflection symmetry implies
\begin{equation}
c(u_1,\dots,u_d)= c(1-u_1,\dots,1-u_d), \quad \mathbf{u} \in [0,1]^d.
\end{equation}
The simplest one-parameter bivariate copula families are exchangeable but not necessarily reflection symmetric. For practical situations these assumptions on copulas are too restrictive and they need to be relaxed for flexible dependence modeling. We can see that the GSN copula in (\ref{gsncop1}) is asymmetric in general. The following theorem state that it can be symmetric under certain situations. We continue with the assumption, $p \in (0,1)$.
\begin{th1}\label{theorem1}
A $d$-dimensional $C_{d,GSN}(p,\mathbf{\mu,\Sigma})$ copula is exchangeable if and only if the correlation matrix $\mathbf{\Sigma}$ is exchangeable and $\mu_j = \mu \in \mathcal{R}$ for all $j = 1,\dots,d$. Moreover, it is radially symmetric when $\mu = 0$.
\end{th1}
%% Start the tail dependence properties. %%
\par Lastly we need to discuss the tail dependence of the GSN copula. Tail dependence quantifies the degree of dependence in the joint lower or joint upper tail of a multivariate distribution. Here we consider the bivariate tail dependence only, but there are multivariate extensions to the concept in the literature (see, \citet{jaworski2010copula}). For a bivariate distribution, tail dependence is defined as the limiting probability of exceeding a certain threshold by one margin given that the other margin has already exceeded that threshold. Upper and lower tail dependence coefficients are of interest for a bivariate distribution or copula. Let $(X_1,X_2)^\intercal$ be a bivariate random vector with marginal distribution functions $F_j, j = 1,2$, and copula $C$. Then the coefficients of upper and lower tail dependence are defined as
\begin{equation}\label{lambu1}
\lambda_U(C) = \lim_{u \to 1-} \frac{1 -2u +C(u,u)}{1-u} = \lim_{u \to 1-} P(U_1 > u|U_2 > u) \quad \text{and}
\end{equation}
\begin{equation}\label{lambl1}
\lambda_L(C) = \lim_{u \to 0+} \frac{C(u,u)}{u} = \lim_{u \to 0+} P(U_1 \leq u|U_2 \leq u) \quad \text{respectively}
\end{equation}
where $U_i = F_i(X_i), \; i = \{1,2\}$ provided the above limits exist. The copula $C$ is said to have upper or lower tail dependence if $\lambda_U,\lambda_L \in (0,1]$. If $\lambda_U = 0$ or $\lambda_L = 0$, we say $C$ has no upper or lower tail dependence. For example bivariate Gaussian copula is independent in both upper and lower tail for any correlation parameter $\mathbf{\rho} \in (-1,1)$ whereas Student-$t$ copula has nonzero symmetric tail dependence as shown by \citet{demarta2005t}.
\par Tail dependence of Azzalini's skew-$t$ copula have been demonstrated by \citet{kollo2017tail}. \citet{fung2016tail} obtained the rate of convergence to $0$, of the tail dependence coefficients for a bivariate skew-normal copula. Geometric skew-normal copula shows similar tail behavior as described in the following Theorem.
\begin{th1}\label{theorem2}
A bivariate GSN copula has neither lower nor upper tail dependence i.e.
\begin{equation}
\lambda_U(C_{GSN}) = \lambda_L(C_{GSN}) = 0,
\end{equation}
for all $\mu_1,\mu_2 \in \mathcal{R}, p \in (0,1)$ and the correlation parameter $\mathbf{\rho} \in (-1,1)$.
\end{th1}
Theorem \ref{theorem1} and \ref{theorem2} reveals that the asymmetric nature of a GSN copula mainly comes from the location parameter $\mathbf{\mu}$ of the underlying normal distributions. The fact that it has zero tail dependence can be argued from \citet{beare2010copulas} also since it the copula is generated from a geometric mixture. For the lack of tail dependence one can think of a skew-$t$ extension of this copula involving an additional degrees of freedom parameter. All proofs regarding the theoretical properties discussed in this section are provided in Appendix \ref{app3}.
%% Section 4. %%
\section{Maximum likelihood estimation}\label{sec4}
Let us assume that the data have been transformed into $m$ independent vector valued observations, $\mathbf{u}_i \in [0,1]^d, i = 1,\dots,m$ using some parametric or non-parametric distribution function. Then the set of observations $\{\mathbf{u}_1,\dots,\mathbf{u}_m\}$ is called a `pseudo sample', which provide only the `dependence' information of the data. The log-likelihood function for the copula parameters based on a sample $\mathcal{U} = \{\mathbf{u}_1,\dots,\mathbf{u}_m\}$ of size $m$ from $C_{d,GSN}(p,\mathbf{\mu},\mathbf{\Sigma})$, using (\ref{copden2}) is given as
\begin{multline}\label{like1}
l(p,\mathbf{\mu},\mathbf{\Sigma}|\mathbf{u}_1,\dots,\mathbf{u}_m) = \sum_{i=1}^m l_i(p,\mathbf{\mu},\mathbf{\Sigma}|\mathbf{u}_i) \\ = \sum_{i=1}^m\Big[\log\Big(\sum_{k=1}^\infty \frac{p(1-p)^{k-1}}{(2\pi k)^{d/2}\sqrt{|\mathbf{\Sigma}|}} e^{-\frac{1}{2k}(\mathbf{x}_i - k\mathbf{\mu})^\intercal \mathbf{\Sigma}^{-1} (\mathbf{x}_i - k\mathbf{\mu})}\Big) - \sum_{j=1}^d \log\Big(\sum_{k=1}^\infty \frac{p(1-p)^{k-1}}{\sqrt{2\pi k}} e^{-\frac{1}{2k}(x_{ij} - k\mu_j)^2}\Big)\Big],
\end{multline}
where $x_{ij} = F_1^{-1}(u_{ij}|\mu_j,1,p)$ denotes the corresponding quantiles. The maximum likelihood estimators (MLEs) can be obtained by maximizing (\ref{like1}) with respect to the unknown parameters. The main advantage of the GSN copula lies in its efficiency in the parameter estimation as the quantile function is independent of the correlation matrix $\mathbf{\Sigma}$. Direct maximization of (\ref{like1}) is quite a challenging issue since it involves solving a $(d + 1 + d(d+1)/2)$ dimensional optimization problem. The problem becomes more severe when $d$ is large. \citet{kundu2017multivariate} pointed out that the density function of a geometric skew-normal distribution can be multimodal and so is the log-likelihood. He provided an EM algorithm to maximize the log-likelihood of an MGSN distribution but that is not applicable for the GSN copula. The effective solution to the maximization problem here is to decompose it into simpler sub-problems.
\par \citet{grippof1999globally} introduced globally convergent block-coordinate techniques for unconstrained optimization. The main idea is to decompose the complicated optimization problem into two simpler estimation sub-problems. Instead of estimating all the parameters simultaneously from the objective function, one can partition the set of parameters into two disjoint blocks in such a way that one block can be optimised at a time, while keeping the other block fixed at some values. The authors showed under sufficient convergence criteria, expressed in terms of conditions on the elementary operations performed on each block component, and of suitable (sequential or parallel) connection rules, two-block decomposition algorithm is globally convergent towards stationary points, even in the absence of convexity or uniqueness assumptions. Since the quantile functions in the GSN copula only involves the parameters $\{p,\mathbf{\mu}\}$, two block-coordinate ascent algorithm provides with very efficient estimation of the parameters. It doesn't require additional restrictions to ensure the positive definiteness of the correlation matrix $\mathbf{\Sigma}$. Also the quantile function can be computed relatively fast though it does not have any analytical expression, with the infinite sum in the distribution function is approximated upto some finite values. Here we apply the Newton's method to obtain the quantiles of the GSN copula. Here the parameters involved in (\ref{like1}) are partitioned into $\mathbf{\theta} = \{\mathbf{\theta}_1,\mathbf{\theta}_2\}$, where $\mathbf{\theta}_1 = \{p,\mathbf{\mu}\}$ and $\mathbf{\theta}_2 = \mathbf{\Sigma}$. Starting with some initial approximations, the following algorithm iteratively updates the parameters over one of the blocks by maximizing (\ref{like1}), while keeping the other block fixed at their current values.
%%%% Second Algorithm. %%%%
\begin{alg1}\label{algo2}
(Two block-coordinate ascent algorithm for the GSN copula)
\begin{itemize}
    \item Step 1: Start with some initial approximations of $\hat{\mathbf{\theta}}_1^0$ and $\hat{\mathbf{\theta}}_2^0$.
    \item Step 2: At the $r$-th iteration, update the estimate $\hat{\mathbf{\theta}}_1^r$ by maximizing $l(\mathbf{\theta}_1,\mathbf{\theta}_2)$ over $\mathbf{\theta}_1$ when $\mathbf{\theta}_2$ is fixed at $\hat{\mathbf{\theta}}_2^{r-1}$, i.e.
    \begin{equation}
    \hat{\mathbf{\theta}}_1^r := \underset{\mathbf{\theta}_1}{\arg\max} \{l(\mathbf{\theta}_1,\hat{\mathbf{\theta}}_2^{r-1})\}. \nonumber
    \end{equation}
    \item Step 3: At the $r$-th iteration, update the estimate $\hat{\mathbf{\theta}}_2^r$ by maximizing $l(\mathbf{\theta}_1,\mathbf{\theta}_2)$ over $\mathbf{\theta}_2$ when $\mathbf{\theta}_1$ is fixed at $\hat{\mathbf{\theta}}_1^r$, i.e.
    \begin{equation}
    \hat{\mathbf{\theta}}_2^r := \underset{\mathbf{\theta}_2}{\arg\max} \{l(\hat{\mathbf{\theta}}_1^r,\mathbf{\theta}_2)\}. \nonumber
    \end{equation}
    \item Step 4: Repeat Steps 2 and 3 until the algorithm converges.
\end{itemize}
\end{alg1}
%%%% %%%%
Standard numerical optimization method with box constraints, L-BFGS-B can be used to find the maximum likelihood estimates in Steps $2$ and $3$ with bounds for the parameter $p$ as $(0,1)$ and for the correlation parameters $\{\rho_{ij};1 \leq i < j \leq d\}$ as $(-1,1)$, respectively. Algorithm \ref{algo2} is also applicable to estimate the parameters from the log-likelihood of MGSN distribution. Note that the algorithm do not guarantee a global maximum, it can reach into a local maximum al well, same as for the EM algorithm. The initial approximations of $\hat{\mathbf{\theta}}_1^0$ and $\hat{\mathbf{\theta}}_2^0$ can be chosen by the combination of method of moments estimation and profile likelihood based on the observed data which results in faster convergence of the proposed algorithm. For a fixed value of $p$, the MOM estimates of $\mathbf{\mu}$ and $\mathbf{\Sigma}$ are given as
\begin{equation}
\tilde{\mathbf{\mu}} = p\bar{\mathbf{X}} \;\; \text{and} \;\;  \tilde{\mathbf{\Sigma}} = p\mathbf{S_X} - p(1-p)\bar{\mathbf{X}}\bar{\mathbf{X}}^\intercal.
\end{equation}
Therefore, the MLE of $p$, denoted by $\tilde{p}$ can be obtained by maximizing the profile log-likelihood function of MGSN distribution with known $\mathbf{\mu}$ and $\mathbf{\Sigma}$, i.e. $l(p,\tilde{\mathbf{\mu}},\tilde{\mathbf{\Sigma}})$, with respect to $p$. Finally the initial estimates of $\mathbf{\mu}$ and $\mathbf{\Sigma}$ become
\begin{equation}
\tilde{\mathbf{\mu}} = \tilde{\mathbf{\mu}}(\tilde{p}) \;\; \text{and} \;\; \tilde{\mathbf{\Sigma}} = \tilde{\mathbf{\Sigma}}(\tilde{p}), \;\; \text{respectively}. 
\end{equation}
The above algorithm is applicable for the general structure of GSN copula for moderate to high dimensions, which is verified in one of our simulation study with unrestricted $\mathbf{\mu}$ and $\mathbf{\Sigma}$. In this setting, the observed information matrix for (\ref{like1}) can be numerically obtained as
\begin{equation}\label{fish1}
I_m(\mathbf{\theta}) := \sum_{i=1}^m \frac{\partial}{\partial\mathbf{\theta}} l_i(\mathbf{\theta|\mathbf{u}_i}) \frac{\partial}{\partial\mathbf{\theta}^\intercal} l_i(\mathbf{\theta}|\mathbf{u}_i),
\end{equation}
which can be used to get the standard errors of the parameter estimates.
%% The next section. %%
\section{Regression models for longitudinal data}\label{sec5}
In longitudinal studies, repeated measurements are collected over time to assess the evolution of the responses with respect to some covariates. Suppose $\mathbf{Y}_i = (Y_{i1},\dots,Y_{in_i})^\intercal$ be a vector of $n_i$ dependent responses for $i$-th subject. The marginal cumulative distribution of a single variable $Y_{ij}$ is denoted by $F(Y_{ij}|\mathbf{x}_{ij},\mathbf{\beta})$ and depends on a $p$-dimensional vector of covariates $\mathbf{x}_{ij}$ and a regression parameter $\mathbf{\beta}$ (for lack of notations we use $p$ here for the number of regression parameters). Our scientific objective is evaluating how the distribution of $Y_{ij}$ varies according to the changes in a vector of $p$ covariates $\mathbf{x}_{ij}$ as well as the dependence among $\mathbf{Y}_i$. When $\mathbf{Y}_i$ is continuous, we consider the marginals to follow a generalized linear model as
\begin{equation}\label{glm1}
g(E(Y_{ij}|\mathbf{x}_{ij})) = \mathbf{x}_{ij}\mathbf{\beta}, \quad j = 1,\dots,n_i,
\end{equation}
where $g(.)$ is a suitable link function and $\mathbf{\beta}$ is a $p\times 1$ vector of regression coefficients. However, under the copula framework any kind of marginals can be used other than those belonging to the exponential family. Then the joint distribution function of $\mathbf{Y}_i$ given $\mathbf{x}_i$ can be expressed as
\begin{equation}\label{copglm1}
F_{n_i}(y_{i1},\dots,y_{in_i}|\mathbf{x}_i) = C_{n_i}(F(y_{i1}|\mathbf{x}_{i1}),\dots,F(y_{in_i}|\mathbf{x}_{in_i})|\mathbf{\phi}_i),
\end{equation}
where $C_{n_i}(.|\mathbf{\phi}_i)$ is a $n_i$-dimensional copula with parameter vector $\mathbf{\phi}_i$. The corresponding density function is given by
\begin{equation}\label{copglm2}
f_{n_i}(y_{i1},\dots,y_{in_i}|\mathbf{x}_i) = c_{n_i}(u_{i1},\dots,u_{in_i}|\mathbf{\phi}_i)\prod_{j=1}^{n_i} f(y_{ij}|\mathbf{x}_{ij}),
\end{equation}
where $u_{ij} = F(y_{ij}|\mathbf{x}_{ij})$. The copula identifies a regression model constructed in way to (i) preserve the marginal univariate distributions and (ii) have separate dependence structure. For a fixed set of marginals different multivariate models can be constructed by various choices of the copula function. Based on $m$ independent observations, the log-likelihood of (\ref{copglm1}) is given as
\begin{equation}\label{copllk1}
l(\mathbf{\beta},\mathbf{\phi}|\mathbf{y},\mathbf{x}) = \sum_{i=1}^m \log c_{n_i}(u_{i1},\dots,u_{in_i}|\mathbf{\phi}_i) + \sum_{i=1}^m \sum_{j=1}^{n_i} \log f(y_{ij}|\mathbf{x}_{ij}).
\end{equation}
\par Since the copula can be uniquely identified for continuous dependent random variables, but that is not the case with discrete variables. Hence latent variable formulation (\citet{agresti2010analysis}) can be used to construct ordered probit models for ordinal data. Let $Y_{ij}$ represent a categorical response with $K$ possible ordered categories and let $Z_{ij}$ be a normally distributed latent variable underneath $Y_{ij}$. Let $\gamma(k),\; 1 < k < K - 1$, be ordered thresholds such that: $-\infty = \gamma(0) < \gamma(1) < \dots < \gamma(K-1) < \gamma(K) = \infty$. Then the ordinal variable have the stochastic representation as
\begin{equation}\label{latord1}
Y_{ij} = k \;\; \text{if} \;\; \gamma(k-1) \leq Z_{ij} < \gamma(k), \;\; k \in \{1,\dots,K\}.
\end{equation}
The threshold parameters can be fixed or freely estimated based on specification of the model. Note that the monotonic increasing nature of the thresholds accounts for the ordered nature of the observed outcomes. We model the latent variable $Z_{ij}$, based on covariate vector $\mathbf{x}_{ij}$ as
\begin{equation}\label{latord2}
Z_{ij}|\mathbf{x}_{ij} = \mathbf{x}_{ij}\mathbf{\beta} + \epsilon_{ij}, \;\; j = 1,\dots,n_i,
\end{equation}
where $\mathbf{\beta}$ is a vector of regression coefficients and $\epsilon_{ij}$ is the error term. To ensure the identifiability of the model we assume $\epsilon_{ij} \sim N(0,1)$ and the intercept of $\mathbf{\beta}$ equal to zero. Therefore the dependence structure of the observed response vector $\mathbf{Y}_i$ is explained through the dependence of the latent vector $\mathbf{Z}_i$ (see, \citet{bhat2010comparison}). Then the joint probability mass function can be written as
\begin{multline}\label{copord1}
P(Y_{i1} = y_{i1},\dots,Y_{in_i} = y_{in_i}|\mathbf{x}_i) = P(\gamma(y_{i1}-1) \leq Z_{i1} < \gamma(y_{i1}),\dots,\gamma(y_{in_i}-1) \leq Z_{in_i} < \gamma(y_{in_i})) \\ = \int_{\gamma(y_{i1}-1)}^{\gamma(y_{i1})} \dots \int_{\gamma(y_{in_i}-1)}^{\gamma(y_{n_i})}c_{n_i}(F(z_{i1}|\mathbf{x}_{i1}),\dots,F(z_{in_i}|\mathbf{x}_{in_i})|\mathbf{\phi}_i)\prod_{j=1}^{n_i} f(z_{ij}|\mathbf{x}_{ij}) dz_{i1}\dots dz_{in_i}.
\end{multline}
The joint PMF in (\ref{copord1}) involves $n_i$-dimensional integral which can be obtained as a finite difference of the CDF of $\mathbf{Z}_i$ as acknowledged by \citet{peter2007correlated} and \citet{madsen2011joint}. But as the dimension $n_i$ increase, evaluating the rectangular probability becomes computationally infeasible as one need to consider summation of $2n_i$ many terms (\citet{nikoloulopoulos2009modeling}). 
\par To circumvent the computational issues with discrete Gaussian copula regression models composite likelihood methods (CML) are often employed (see, \citet{varin2010mixed} and \citet{varin2011overview}). These pseudo-likelihood methods are useful when all the multivariate parameters can be identified from lower dimensional marginals. Similar to multivariate Gaussian or Student-$t$ copula, our proposed GSN copula permits to construct composite likelihood combining likelihoods for pairs of observation, because of the description of the quantile function in (\ref{gsncop1}). The pair-wise likelihood approach that we implement here, involves only 2-dimensional integrals. This is a major computational advantage, if compared with the skew-elliptical copulas derived from \citet{azzalini1996multivariate}. Based on $m$ independent observations the pairwise log-likelihood can be written as
\begin{multline}\label{copllk2}
l_c(\mathbf{\beta},\mathbf{\phi}|\mathbf{y},\mathbf{x}) = \sum_{i=1}^m \sum_{j=1}^{n_i-1} \sum_{k = j+1}^{n_i} \log P(Y_{ij} = y_{ij},Y_{ik} = y_{ik}|\mathbf{x}_{i(j,k)}) \\ = \sum_{i=1}^m \sum_{j=1}^{n_i-1} \sum_{k = j+1}^{n_i} \log \Big[C_2(u_{ij},u_{ik}|\mathbf{\phi}_{jk}) - C_2(u^-_{ij},u_{ik} |\mathbf{\phi}_{jk}) - C_2(u_{ij},u^-_{ik}|\mathbf{\phi}_{jk}) + C_2(u^-_{ij},u^-_{ik}|\mathbf{\phi}_{jk})\Big],
\end{multline}
where $u_{il} = \Phi(\gamma(y_{il}) - \mathbf{x}_{il}\mathbf{\beta})$ and $u^-_{il} = \Phi(\gamma(y_{il}-1) - \mathbf{x}_{il}\mathbf{\beta})$, for $l = j,k$ respectively. We use geometric skew-normal copula to construct flexible dependence models for continuous and discrete longitudinal data. \citet{masarotto2017gaussian} discussed Gaussian copula regression models along with their computational implementations in R. In order to account for the within-subject dependency or temporal dependency, appropriate structures for the correlation matrix $\mathbf{\Sigma}$ can be considered in our GSN copula based regression models. In particular, we implement the following structures as
\begin{itemize}\label{str1}
    \item Exchangeable ($EX$): $\rho_{jk} = \rho \in (-1,1), \quad 1 \leq j < k \leq n_i$;
    \item $AR(1)$ with exponential decay : $\rho_{jk} = \exp(-\xi|t_j - t_k|), \xi > 0, \quad 1 \leq j < k \leq n_i$.
\end{itemize}
Additionally, we assume equal value of the parameter $\mathbf{\mu}$ for each dimension, i.e. $\mu_j = \mu, \;\; j = 1,\dots,n_i$. That leads to reduction in the number of estimable parameters from the model, for moderate to high dimensional longitudinal data. The parameters of the considered regression models can be obtained from the likelihood and pseudo-likelihood functions in (\ref{copllk1}) and (\ref{copllk2}). But under the regression setup, direct maximization of these is still computationally challenging, especially for complex dependence structure of the GSN copula, since we have additional marginal parameters in the regression models. To obtain valid parameter estimates, we employ the two-stage estimation method often known as inference function for margins (IFM) by \citet{joe1996estimation} and \citet{joe2005asymptotic}. Under this, we estimate the marginal parameters, say $\mathbf{\theta}$, from marginal likelihoods assuming independence. Then at the second step, the copula parameters, say $\mathbf{\phi}$, are estimated from the multivariate likelihood or the composite likelihood with univariate parameter estimates held fixed. The standard errors of the parameter estimates $\hat{\mathbf{\theta}}^* = (\hat{\mathbf{\theta}},\hat{\mathbf{\phi}})^\intercal$ can be numerically obtained from the observed sandwich information matrix (Godambe information matrix) as
\begin{equation}\label{godam1}
J(\hat{\mathbf{\theta}}^*) = D(\hat{\mathbf{\theta}}^*)^\intercal M(\hat{\mathbf{\theta}}^*)^{-1} D(\hat{\mathbf{\theta}}^*),
\end{equation}
where $D(\hat{\mathbf{\theta}}^*)$ is a block diagonal matrix and $M(\hat{\mathbf{\theta}}^*)$ is a symmetric positive definite matrix. The explicit forms of these can be found in \citet{zhao2005composite} or \citet{joe2014dependence}. To estimate the parameters we use {\em optim} function, and to estimate the information matrix associted with the parameter estimates we use {\em numderiv} function in R.
%% End of this section. %%
\section{Model comparison}\label{sec6}
For our proposed GSN copula based regression models we wish to compare the fits with the corresponding Gaussian copula based regression models with the same structure of the correlation matrix and investigate for improvements, if any. For this purpose we consider Akaike information criterion and one of its modified version, evaluated at the parameter estimates $\hat{\mathbf{\theta}}^*$ as
\begin{itemize}
    \item AIC under correct specification of the copula and the marginals by \citet{ko2019copula}:
    \begin{equation}
    AIC = -2l(\hat{\mathbf{\theta}}^*) + 2\dim(\hat{\mathbf{\theta}}^*),
    \end{equation}
    \item Composite likelihood version of AIC, as given in \citet{varin2005note}:
    \begin{equation}
    CLAIC = -2l_c(\hat{\mathbf{\theta}}^*) + 2tr(M(\hat{\mathbf{\theta}}^*)D(\hat{\mathbf{\theta}}^*)^{-1}),
    \end{equation}
\end{itemize}
for the continuous and ordinal regression models respectively. The smaller values of these criteria leads to better fitting regression models. Note that the matrices used in CLAIC is not same as in \ref{godam1}. But we use the exact form of CLAIC as mentioned in the previously cited paper by assuming the two step estimates are very close to the estimates if one step estimation of the composite likelihood was obtained. But with this we can compare different models as will be described next.
\par In addition, we will also use Voung's test (\citet{vuong1989likelihood}) to show if a GSN copula based model provides a better fit than Gaussian copula model with same structure of the correlation matrix. Voung's test is the sample version of the difference in Kullback-Leibler divergence and sample size to differentiate two models which could be non-nested. This test has been used extensively in the copula literature to compare vine copula models (e.g., \citet{brechmann2012truncated}; \citet{joe2014dependence} or \citet{nikoloulopoulos2017vine}). Here we provide the details in a general context. Assume that we have two models $M_1$ and $M_2$, with parametric densities $f^{(1)}_\mathbf{y}$ and $f^{(2)}_\mathbf{y}$ respectively, we can compare
\begin{align}\label{voung1}
\Delta_{1f} & = \frac{1}{m}\Big[\sum_{i=1}^m \Big\{ E_f\log f_\mathbf{y}(\mathbf{y}_i) - E_f\log f^{(1)}_\mathbf{y}(\mathbf{y}_i|\mathbf{\theta}^*_1) \Big\}\Big], \nonumber \\ \Delta_{2f} & = \frac{1}{m}\Big[\sum_{i=1}^m \Big\{ E_f\log f_\mathbf{y}(\mathbf{y}_i) - E_f\log f^{(2)}_\mathbf{y}(\mathbf{y}_i|\mathbf{\theta}^*_2) \Big\}\Big],
\end{align}
where $\mathbf{\theta}_1,\mathbf{\theta}_2$ are the parameters in models $M_1$ and $M_2$, respectively, that lead to the closest Kullback-Leibler divergence to the true $f_\mathbf{y}$; equivalently, they are the limits in probability of the ML estimates based on models $M_1$ and $M_2$, respectively. Model $M_1$ is closer to the true $f_{\mathbf{y}}$, i.e., it is the better-fitting model if $\Delta_{12} = \Delta_{1f} - \Delta_{2f} < 0$, and Model $M_2$ is the better-fitting model if $\Delta_{12} > 0$. The sample version of $\Delta_{12}$ with ML estimates $\hat{\mathbf{\theta}}^*_1,\hat{\mathbf{\theta}}^*_2$ is
\begin{equation}\label{voung2}
\bar{D}_{12} = \frac{1}{m} \sum_{i=1}^m D_i, \;\; \text{where} \;\; D_i = \log\frac{f^{(2)}_\mathbf{y}(\mathbf{y}_i|\hat{\mathbf{\theta}}^*_2)}{f^{(1)}_\mathbf{y}(\mathbf{y}_i|\hat{\mathbf{\theta}}^*_1)}.
\end{equation}
In our setup we use two stage estimates assuming they are very close to the true ML estimates. For non-nested or nested models where $f^{(1)}_\mathbf{y}(\mathbf{y}_i|\mathbf{\theta}^*_1)$ and $f^{(2)}_\mathbf{y}(\mathbf{y}_i|\mathbf{\theta}^*_2)$ are not the same density, a large sample $95\%$ confidence interval (CI) for the parameter $\Delta_{12}$ is
\begin{equation}\label{voungci}
\bar{D}_{12} \pm 1.96 \times \frac{\bar{s}_{12}}{\sqrt{m}}, \;\; \text{where} \;\; \bar{s}_{12} = \frac{1}{m-1}\sum_{i=1}^m (D_i - \bar{D}_{12})^2.
\end{equation}
If the interval in (\ref{voungci}) contains $0$, models $M_1$ and $M_2$ would not be considered significantly different, which can be used as a diagnostic in our comparison. Point to be noted that, Voung's test is applicable for both likelihood and pseudo-likelihood based methods, but the condition is the expectations defined in (\ref{voung1}), should consistently estimate the model parameters. However, Voung's test does not evaluate whether any of the models provide a sufficiently good fit, specifically in terms of how closely the model approximates the true data-generating mechanism.
%% Section 6. %%
\section{Simulation studies}\label{sec7}
In this section we investigate the finite sample performance of the parametric inference of the proposed GSN copula and the considered regression models in Section \ref{sec5}. We consider $3$ simulation studies. We generate random data sets form the respective models and then estimate the parameters using the methods described in Section \ref{sec4} and \ref{sec5}. For each simulations we consider $2$ different sample sizes as $m = \{200,500\}$ and for the first simulation we consider the number of samples to be $200$ and in the subsequent $2$ simulations we consider the replication number to be $500$.
\par For the unrestricted case of the MGSN distribution and the GSN copula we consider the following set of parameters -
\begin{equation}\label{sim1}
p = 0.5, \quad \mathbf{\mu} = \left(\begin{matrix} 0 \\ 0 \\ 1 \\ 1 \end{matrix}\right), \quad \mathbf{\Sigma} = \left(\begin{matrix} 1 & 0.6 & 0.4 & 0.2 \\ 0.6 & 1 & 0.2 & 0.4 \\ 0.4 & 0.2 & 1 & 0.2 \\ 0.2 & 0.4 & 0.2 & 1 \end{matrix}\right).
\end{equation}
The data sets, say $\mathcal{X}$ (MGSN distribution) and $\mathcal{U}$ (GSN copula) of size $m$ are generated using Algorithm \ref{algo1}. Then we calculate the MLEs of the parameters using Algorithm \ref{algo2}, for the distribution and the copula data, respectively. The maximization of the log-likelihood of the GSN copula takes significantly more time than the log-likelihood of the MGSN distribution, since it involves computation of the quantiles in each block. Two block-coordinate ascent algorithm converges within $5$ iterations for the MGSN distribution and $15$ iterations for the GSN copula respectively. The starting values of the parameters are obtained by the methods discussed in Section $4$.
%% The first table. %%
\begin{table}
    \centering
    \begin{small}
    \rotatebox{90}{
    \begin{minipage}{0.98\textwidth}
    \scalebox{1.0}{
    \tabcolsep = 0.18cm
    \begin{tabular}{|c c|c c c c c|c c c c c|}
    \hline
    \multicolumn{2}{|c|}{} & \multicolumn{5}{c|}{\textbf{m} = 200} & \multicolumn{5}{c|}{\textbf{m} = 500} \\
    \hline
    \textbf{Parameters} & \textbf{True Value} & Mean & Bias & SD & SE & RMSE & Mean & Bias & SD & SE & RMSE \\
    \hline
    \multicolumn{2}{|c}{\textbf{MGSN Distribution}} & \multicolumn{10}{c|}{} \\
    \hline
    $p$ & 0.5 & 0.5172 & 0.0172 & 0.0676 & 0.0681 & 0.0698 & 0.5092 & 0.0092 & 0.0495 & 0.0491 & 0.0504 \\
    $\mu_1$ & 0.0 & -0.0035 & -0.0035 & 0.0526 & 0.0525 & 0.0527 & 0.0027 & 0.0027 & 0.0289 & 0.0287 & 0.0289 \\
    $\mu_2$ & 0.0 & -0.0060 & -0.0060 & 0.0534 & 0.0536 & 0.0538 & 0.0035 & 0.0035 & 0.0331 & 0.0330 & 0.0333 \\
    $\mu_3$ & 1.0 & 1.0285 & 0.0285 & 0.1376 & 0.1380 & 0.1405 & 1.0170 & 0.0170 & 0.0996 & 0.0990 & 0.1011 \\
    $\mu_4$ & 1.0 & 1.0254 & 0.0254 & 0.1421 & 0.1432 & 0.1443 & 1.0136 & 0.0136 & 0.1043 & 0.1041 & 0.1052 \\
    $\rho_{12}$ & 0.6 & 0.5955 & -0.0045 & 0.0444 & 0.0446 & 0.0446 & 0.5945 & -0.0055 & 0.0248 & 0.0247 & 0.0252 \\
    $\rho_{13}$ & 0.4 & 0.3895 & -0.0105 & 0.0801 & 0.0818 & 0.0833 & 0.3928 & -0.0072 & 0.0544 & 0.0538 & 0.0561 \\
    $\rho_{14}$ & 0.2 & 0.1932 & -0.0068 & 0.0998 & 0.1008 & 0.1011 & 0.1937 & -0.0063 & 0.0606 & 0.0605 & 0.0610 \\
    $\rho_{23}$ & 0.2 & 0.1947 & -0.0053 & 0.1106 & 0.1117 & 0.1122 & 0.1950 & -0.0050 & 0.0578 & 0.0572 & 0.0581 \\
    $\rho_{24}$ & 0.4 & 0.3963 & -0.0037 & 0.0886 & 0.0888 & 0.0891 & 0.3965 & -0.0035 & 0.0605 & 0.0602 & 0.0608 \\
    $\rho_{34}$ & 0.2 & 0.2179 & 0.0179 & 0.0943 & 0.0950 & 0.0956 & 0.2123 & 0.0123 & 0.0667 & 0.0663 & 0.0679 \\
    \hline
    \multicolumn{2}{|c}{\textbf{GSN Copula}} & \multicolumn{10}{c|}{} \\
    \hline
    $p$ & 0.5 & 0.5266 & 0.0266 & 0.1051 & 0.1062 & 0.1084 & 0.5019 & 0.0019 & 0.0630 & 0.0628 & 0.0631 \\
    $\mu_1$ & 0.0 & -0.0272 & -0.0272 & 0.1325 & 0.1333 & 0.1353 & -0.0036 & -0.0036 & 0.0767 & 0.0762 & 0.0768 \\
    $\mu_2$ & 0.0 & -0.0285 & -0.0285 & 0.1482 & 0.1490 & 0.1509 & 0.0074 & 0.0074 & 0.0751 & 0.0749 & 0.0754 \\
    $\mu_3$ & 1.0 & 1.1805 & 0.1805 & 0.4956 & 0.5117 & 0.5274 & 1.0697 & 0.0697 & 0.2124 & 0.2112 & 0.2235 \\
    $\mu_4$ & 1.0 & 1.0704 & 0.0704 & 0.3294 & 0.3312 & 0.3368 & 1.0550 & 0.0550 & 0.2098 & 0.2090 & 0.2169 \\
    $\rho_{12}$ & 0.6 & 0.5992 & -0.0008 & 0.0477 & 0.0482 & 0.0478 & 0.5976 & -0.0024 & 0.0277 & 0.0275 & 0.0286 \\
    $\rho_{13}$ & 0.4 & 0.4133 & 0.0133 & 0.0904 & 0.0911 & 0.0914 & 0.4063 & 0.0063 & 0.0612 & 0.0610 & 0.0615 \\
    $\rho_{14}$ & 0.2 & 0.2144 & 0.0144 & 0.1059 & 0.1063 & 0.1069 & 0.1979 & -0.0021 & 0.0694 & 0.0690 & 0.0697 \\
    $\rho_{23}$ & 0.2 & 0.2004 & 0.0004 & 0.1253 & 0.1258 & 0.1253 & 0.1974 & -0.0026 & 0.0600 & 0.0599 & 0.0604 \\
    $\rho_{24}$ & 0.4 & 0.4106 & 0.0106 & 0.0909 & 0.0912 & 0.0915 & 0.3976 & -0.0024 & 0.0639 & 0.0632 & 0.0642 \\
    $\rho_{34}$ & 0.2 & 0.1907 & -0.0093 & 0.1189 & 0.1192 & 0.1193 & 0.1893 & -0.0107 & 0.0826 & 0.0821 & 0.0841 \\
    \hline
    \end{tabular}}
    \caption{Parameter estimation for multivariate geometric skew-normal distribution and geometric skew-normal copula for $N = 200$ simulated data sets with two different sample sizes.}
    \label{tab:simulation1}
    \end{minipage}}
    \end{small}
\end{table}
%% Continuous data. %%
\par Next we consider regression models for continuous data with generalized linear model for the marginals including a continuous time-varying covariate and GSN copula. We take structured correlation matrix $\mathbf{\Sigma}$ and equal value of $\mathbf{\mu}$ denoted as $\bar{\mu}$ for the GSN copula. The response distribution for all the marginals are taken as Gamma with $\log$ link function. First we sample the copula data and then use PIT to generate response variables from the model -
\begin{equation}\label{simglm1}
g(E(Y_{ij})) = \beta_0 + x_{i1}\beta_1 + x_{i2}\beta_2 + t_{ij}\beta_3, \quad j = 1,\dots,4,
\end{equation}
where the response distribution is Gamma ($\log$-link). For values of the marginal parameters we set $\beta_0 = 1.0$, $\beta_1 = 0.5$, $\beta_2 = 0.5$, $\beta_3 = 1.0$ and the shape parameter $\kappa = 3$. The covariates are generated as $x_{i1} \sim Ber(p = 0.5)$, $x_{i2} \sim N(5,4)$ and the time points $t_{ij} = j$ for $j = 1,\dots,4$. We consider exchangeable and $AR(1)$ correlation structure for the matrix $\mathbf{\Sigma}$, and in both the scenarios we set the autocorrelation parameter $\xi = 0.50$. Finally for other two parameters of the reduced GSN copula we set $p = 0.5$ and $\bar{\mu} = 1.0$, respectively.
%% The second table. %%
\begin{table}[]
    \centering
    \begin{small}
    \rotatebox{90}{
    \begin{minipage}{0.98\textwidth}
    \scalebox{1.0}{
    \tabcolsep = 0.18cm
    \begin{tabular}{|c c|c c c c c|c c c c c|}
    \hline
    \multicolumn{2}{|c|}{} & \multicolumn{5}{c|}{\textbf{m} = 200} & \multicolumn{5}{c|}{\textbf{m} = 500} \\
    \hline
    \textbf{Parameters} & \textbf{True Value} & Mean & Bias & SD & SE & RMSE & Mean & Bias & SD & SE & RMSE \\
    \hline
    \multicolumn{2}{|c}{\textbf{Exchangeable}} & \multicolumn{10}{c|}{} \\
    \hline
    $\beta_0$ & 1.0 & 1.0163 & 0.0163 & 0.1114 & 0.1079 & 0.1126 & 1.0144 & 0.0144 & 0.0744 & 0.0692 & 0.0750 \\
    $\beta_1$ & 0.5 & 0.4995 & -0.0005 & 0.0752 & 0.0744 & 0.0752 & 0.5015 & 0.0015 & 0.0487 & 0.0473 & 0.0487 \\
    $\beta_2$ & 0.5 & 0.4974 & -0.0026 & 0.0188 & 0.0185 & 0.0189 & 0.4968 & -0.0032 & 0.0120 & 0.0118 & 0.0124 \\
    $\beta_3$ & 1.0 & 0.9977 & -0.0023 & 0.0096 & 0.0082 & 0.0099 & 0.9983 & -0.0017 & 0.0060 & 0.0052 & 0.0062 \\
    $\kappa$ & 3.0 & 2.9994 & -0.0006 & 0.2380 & 0.2227 & 0.2380 & 2.9962 & -0.0038 & 0.1630 & 0.1411 & 0.1637 \\
    $p$ & 0.5 & 0.5095 & 0.0095 & 0.1236 & 0.1079 & 0.1240 & 0.4967 & -0.0033 & 0.0744 & 0.0646 & 0.0744 \\
    $\xi$ & 0.5 & 0.5113 & 0.0113 & 0.1024 & 0.1001 & 0.1030 & 0.5023 & 0.0023 & 0.0607 & 0.0596 & 0.0607 \\
    $\bar{\mu}$ & 1.0 & 1.0469 & 0.0469 & 0.2962 & 0.2848 & 0.2999 & 1.0081 & 0.0081 & 0.1687 & 0.1697 & 0.1689 \\
    \hline
    \multicolumn{2}{|c}{\textbf{Autoregressive}} & \multicolumn{10}{c|}{} \\
    \hline
    $\beta_0$ & 1.0 & 1.0064 & 0.0064 & 0.1171 & 0.1072 & 0.1173 & 1.0027 & 0.0027 & 0.0724 & 0.0684 & 0.0753 \\
    $\beta_1$ & 0.5 & 0.5041 & 0.0041 & 0.0745 & 0.0718 & 0.0746 & 0.4957 & -0.0043 & 0.0456 & 0.0458 & 0.0458 \\
    $\beta_2$ & 0.5 & 0.4980 & -0.0020 & 0.0199 & 0.0178 & 0.0200 & 0.4976 & -0.0024 & 0.0117 & 0.0114 & 0.0120 \\
    $\beta_3$ & 1.0 & 0.9996 & -0.0004 & 0.0125 & 0.0123 & 0.0125 & 0.9989 & -0.0011 & 0.0086 & 0.0078 & 0.0088 \\
    $\kappa$ & 3.0 & 3.0124 & 0.0124 & 0.2363 & 0.2283 & 0.2366 & 2.9831 & -0.0169 & 0.1483 & 0.1307 & 0.1498 \\
    $p$ & 0.5 & 0.5049 & 0.0049 & 0.1114 & 0.0960 & 0.1115 & 0.5041 & 0.0041 & 0.0803 & 0.0606 & 0.0804 \\
    $\xi$ & 0.5 & 0.5106 & 0.0106 & 0.0810 & 0.0764 & 0.0817 & 0.4968 & -0.0032 & 0.0529 & 0.0468 & 0.0530 \\
    $\bar{\mu}$ & 1.0 & 1.0355 & 0.0355 & 0.2740 & 0.2466 & 0.2763 & 1.0226 & 0.0226 & 0.1833 & 0.1612 & 0.1847 \\
    \hline
    \end{tabular}}
    \caption{Parameter estimation for geometric skew-normal copula model with Gamma marginals for $N = 500$ simulated data sets. Exchangeable and autoregressive correlation structures are considered.}
    \label{tab:simulation2}
    \end{minipage}}
    \end{small}
\end{table}
\par Finally we consider regression models for ordinal data with latent latent variable formulation including a continuous time-varying covariate and the dependence structure is framed through GSN copula. The parametric structure of the GSN copula is same as the previous study. Here we consider the ordered probit model as -
\begin{align}\label{simord1}
Y_{ij} & = k \;\; \text{if} \;\; \gamma(k-1) \leq Z_{ij} < \gamma(k), \;\; k = 1,\dots,4, \nonumber \\
Z_{ij} & = x_{i1}\beta_1 + x_{i2}\beta_2 + t_{ij}\beta_3 + \epsilon_{ij}, \quad j = 1,\dots,4,
\end{align}
where $\epsilon_{ij} (i.i.d) \sim N(0,1)$. Here we set the marginal parameters, $\beta_1 = 0.5$, $\beta_2 = 0.5$, $\beta_3 = 1.0$ and the threshold parameters $\gamma_1 = 2.0$, $\gamma_2 = 4.0$, $\gamma_3 = 6.0$, respectively. The covariates are generated in the similar format of the previous study. First we generate the copula data using same set of parameters as previous, then use (\ref{simord1}) to obtain the ordinal response variables.
%% The third table. %%
\begin{table}
    \centering
    \begin{small}
    \rotatebox{90}{
    \begin{minipage}{0.98\textwidth}
    \scalebox{1.0}{
    \tabcolsep = 0.18cm
    \begin{tabular}{|c c|c c c c c|c c c c c|}
    \hline
    \multicolumn{2}{|c|}{} & \multicolumn{5}{c|}{\textbf{m} = 200} & \multicolumn{5}{c|}{\textbf{m} = 500} \\
    \hline
    \textbf{Parameters} & \textbf{True Value} & Mean & Bias & SD & SE & RMSE & Mean & Bias & SD & SE & RMSE \\
    \hline
    \multicolumn{2}{|c}{\textbf{Exchangeable}} & \multicolumn{10}{c|}{} \\
    \hline
    $\beta_1$ & 0.5 & 0.5048 & 0.0048 & 0.1348 & 0.1388 & 0.1348 & 0.5027 & 0.0027 & 0.0841 & 0.0878 & 0.0841 \\
    $\beta_2$ & 0.5 & 0.5081 & 0.0081 & 0.0423 & 0.0420 & 0.0431 & 0.5030 & 0.0030 & 0.0264 & 0.0264 & 0.0271 \\
    $\beta_3$ & 1.0 & 1.0144 & 0.0144 & 0.0531 & 0.0536 & 0.0550 & 1.0039 & 0.0039 & 0.0354 & 0.0338 & 0.0356 \\
    $\gamma_1$ & 2.0 & 2.0300 & 0.0300 & 0.2563 & 0.2450 & 0.2580 & 2.0149 & 0.0149 & 0.1618 & 0.1550 & 0.1625 \\
    $\gamma_2$ & 4.0 & 4.0600 & 0.0600 & 0.2795 & 0.2853 & 0.2858 & 4.0270 & 0.0270 & 0.1845 & 0.1801 & 0.1865 \\
    $\gamma_3$ & 6.0 & 6.0865 & 0.0865 & 0.3530 & 0.3563 & 0.3634 & 6.0334 & 0.0334 & 0.2344 & 0.2249 & 0.2368 \\
    $p$ & 0.5 & 0.5092 & 0.0092 & 0.1217 & 0.1360 & 0.1229 & 0.5079 & 0.0079 & 0.1033 & 0.1064 & 0.1094 \\
    $\xi$ & 0.5 & 0.5174 & 0.0174 & 0.1762 & 0.1800 & 0.1801 & 0.5041 & 0.0041 & 0.1245 & 0.1283 & 0.1251 \\
    $\bar{\mu}$ & 1.0 & 1.0789 & 0.0789 & 0.4466 & 0.5140 & 0.4662 & 1.0244 & 0.0244 & 0.3587 & 0.3616 & 0.3631 \\
    \hline
    \multicolumn{2}{|c}{\textbf{Autoregressive}} & \multicolumn{10}{c|}{} \\
    \hline
    $\beta_1$ & 0.5 & 0.4909 & -0.0091 & 0.1292 & 0.1335 & 0.1296 & 0.5003 & 0.0003 & 0.0792 & 0.0743 & 0.0792 \\
    $\beta_2$ & 0.5 & 0.5083 & 0.0083 & 0.0412 & 0.0404 & 0.0420 & 0.5015 & 0.0015 & 0.0253 & 0.0225 & 0.0254 \\
    $\beta_3$ & 1.0 & 1.0132 & 0.0132 & 0.0531 & 0.0546 & 0.0547 & 1.0025 & 0.0025 & 0.0344 & 0.0334 & 0.0345 \\
    $\gamma_1$ & 2.0 & 2.0350 & 0.0350 & 0.2358 & 0.2448 & 0.2383 & 2.0063 & 0.0063 & 0.1538 & 0.1536 & 0.1539 \\
    $\gamma_2$ & 4.0 & 4.0615 & 0.0615 & 0.2749 & 0.2808 & 0.2817 & 4.0119 & 0.0119 & 0.1771 & 0.1768 & 0.1775 \\
    $\gamma_3$ & 6.0 & 6.0835 & 0.0825 & 0.3389 & 0.3475 & 0.3490 & 6.0168 & 0.0168 & 0.2190 & 0.2187 & 0.2197 \\
    $p$ & 0.5 & 0.5130 & 0.0130 & 0.1333 & 0.1402 & 0.1365 & 0.5032 & 0.0032 & 0.0983 & 0.0959 & 0.0997 \\
    $\xi$ & 0.5 & 0.5120 & 0.0120 & 0.1642 & 0.1756 & 0.1672 & 0.5049 & 0.0049 & 0.1074 & 0.1062 & 0.1084 \\
    $\bar{\mu}$ & 1.0 & 1.0585 & 0.0585 & 0.4344 & 0.4597 & 0.4642 & 1.0405 & 0.0405 & 0.3349 & 0.3325 & 0.3412 \\
    \hline    
    \end{tabular}}
    \caption{Parameter estimation for geometric skew-normal copula based ordered probit models for $N = 500$ simulated data sets. Exchangeable and autoregressive correlation structures are considered.}
    \label{tab:simulation3}
    \end{minipage}}
    \end{small}
\end{table}
\par Table \ref{tab:simulation1}, \ref{tab:simulation2} and \ref{tab:simulation3} presents the simulation results for the considered models. Here we report the mean, the biases [$\frac{1}{N}\sum_{i=1}^N (\hat{\mathbf{\theta}}^*_j - \mathbf{\theta}^*)$], empirical standard deviations (denoted as SD), average standard errors obtained from the asymptotic covariance matrices (denoted as SE) and roots of mean square errors [$\sqrt{\frac{1}{N}\sum_{i=1}^N (\hat{\mathbf{\theta}}^*_j - \mathbf{\theta}^*)^2}$], where $\hat{\mathbf{\theta}}^*_j$ is the parameter estimates for the $j$-th sample. The mean estimates are close to the true values and the RMSEs go to zero with increase in sample size. From Table \ref{tab:simulation1}, we see that standard errors of the parameters $\{\mu_j; j = 1,\dots,4\}$ are little more when estimated from the GSN copula likelihood, simply because it involves calculation of the quantiles. From Table \ref{tab:simulation2} and \ref{tab:simulation3}, we see that bias and RMSE of the autocorrelation parameter $\xi$, is slightly bigger in the model with $EX$ correlation matrix $\Sigma$, than the model with $AR(1)$ correlation matrix. SE and SD are consistent with each other for all the cases which suggests the methods proposed for the models are valid. Overall, the studies conducted in this section are encouraging enough to apply our models to analyze real life data sets.
%% This section ends here. %%
\section{Applications}\label{sec8}
In this section we illustrate the flexibility of the regression models described in this paper through some examples, and compare the fits with the corresponding Gaussian copula based models. The datasets considered are publicly available in several R packages such as {\em qrLMM} or {\em mixor}.
%% New subsection. %%
\subsection{Framingham heart study}
This is a benchmark data set in longitudinal studies, which was previously analyzed by several authors, such as \citet{zhang2001linear} and \citet{arellano2005skew}. The data set provides cholesterol levels over time, age at baseline and gender for $200$ randomly selected patients, measured at the beginning of the study and every two years for a total of $10$ years. The primary objective is to model the change of cholesterol levels over time within patients. However, we apply KNN method to impute the missing entries in this data set beforehand. Since the cholesterol levels observed to be positively skewed, we consider Gamma distribution ($\log$-link) for the marginals. We adopt the following model -
\begin{equation}\label{fram1}
g(E(Y_{ij})) = \beta_0 + \text{sex}_i\beta_1 + \text{age}_i\beta_2 + t_{ij}\beta_3, \quad j = 1,\dots,6,
\end{equation}
where observed $y_{ij}$ is cholesterol level divided by $100$ at the $j$-th time for subject $i$. The available covariates are: $t_{ij} = (\text{time} - 5)/10$ (time measured in years), sex ($0$ = female, $1$ = male) and age at baseline. We consider GSN and Gaussian copula to model the temporal dependence of the cholesterol levels. The sample correlation matrix of the responses suggests that exchangeable correlation structure for the matrix $\Sigma$ is adequate, which we reparameterize by $\rho = \exp(-\xi), \xi > 0$.
%% The table starts here. %%
\begin{table}
    \centering
    \begin{small}
    \scalebox{1.0}{
    \tabcolsep = 0.18cm
    \begin{tabular}{|c|c c c c c c|}
    \hline
    \multicolumn{1}{|c|}{Marginal} & \multicolumn{6}{c|}{} \\
    \hline
    \textbf{Parameters} & $\beta_0$ & $\beta_1$ & $\beta_2$ & $\beta_3$ & $\kappa$ & \\
    \hline
    Est. & 0.5861 & -0.0061 & 0.0063 & 0.1181 & 33.0399 & \\
    SE & 0.0652 & 0.0225 & 0.0016 & 0.0091 & 2.8327 & \\
    \hline
    \multicolumn{1}{|c|}{Copula} & \multicolumn{3}{c|}{GSN} & \multicolumn{3}{c|}{Gaussian} \\
    \hline
    \multicolumn{1}{|c|}{\textbf{Parameters}} & \multicolumn{1}{c}{$p$} & \multicolumn{1}{c}{$\xi$} & \multicolumn{1}{c|}{$\bar{\mu}$} & \multicolumn{3}{c|}{$\xi$} \\
    \hline
    \multicolumn{1}{|c|}{Est.} & \multicolumn{1}{c}{0.7185} & \multicolumn{1}{c}{0.3250} & \multicolumn{1}{c|}{-0.1577} & \multicolumn{3}{c|}{0.3284} \\
    \multicolumn{1}{|c|}{SE} & \multicolumn{1}{c}{0.0647} & \multicolumn{1}{c}{0.0214} & \multicolumn{1}{c|}{0.1514} & \multicolumn{3}{c|}{0.0172} \\
    \hline
    \multicolumn{1}{|c|}{Likelihood} & \multicolumn{3}{c|}{\textbf{-116.47}} & \multicolumn{3}{c|}{-129.61} \\
    \multicolumn{1}{|c|}{AIC} & \multicolumn{3}{c|}{\textbf{248.95}} & \multicolumn{3}{c|}{271.22} \\
    \hline
    \multicolumn{7}{|c|}{$D_{12}$ = 0.0657, 95\% = (0.0140,0.1174), p-val = \textbf{0.0127}} \\
    \hline
    \end{tabular}}
    \end{small}
    \caption{Fitting of cholesterol data under model (\ref{fram1}) with GSN and Gaussian copula. Observed log-likelihoods, AICs and the summary of Voung's statistic are reported.}
    \label{tab:realcont1}
\end{table}
\par Table \ref{tab:realcont1} presents the parameters estimates, standard errors for model (\ref{fram1}) with the GSN and the Gaussian copula. It also displays the observed log-likelihoods, AICs and the summary of Voung's statistic for comparison of two models. It is evident that GSN copula better explains the temporal dependence of the cholesterol levels, since confidence interval of $D_{12}$ does not contain zero and also the observed AIC is minimum for this model. The value of correlation parameter is very close under both the copulas, which validates the assumed exchangeable correlation structure for $\mathbf{\Sigma}$. The estimates of the regression parameters $\beta_1$ and $\beta_2$ are close to zero, suggesting that patient's gender and age have insignificant effect on the change in cholesterol levels. Moreover, the underlying copula is negatively skewed, as shown by the value of $\bar{\mu}$.
%% The next subsection. %%
\subsection{Schizophrenia collaborative study}
This data set is from the National Institute of Mental Health Schizophrenia Collaborative Study, previously analyzed by \citet{gibbons1988random} or \citet{gibbons1994application}. Patients were randomly assigned to receive one of four medications, either placebo or one of thee different anti-psychotic drugs (chlorpromazine, fluphenazine or thioridazine). Here we analyze the outcome variable {\em imps79o}, which is an ordinally scaled version of the original variable {\em imps79}. This scaling was done in \citet{gibbons1994application} to retain more information about the response but to ensure each response category has a relatively large number of respondents, since some response categories had relatively small number of subjects compared to others. The ordinal response variable has the following interpretation: $1$ = not ill or borderline; $2$ = mildly or moderately ill; $3$ = markedly ill; and $4$ = severely or most extremely ill. Here we perform complete data analysis for $308$ patients who were evaluated at weeks $0,1,3$ and $6$ to assess severity of illness. The covariates are taken as treatment ($0$ = placebo, $1$ = drug) and the square root of the time variable (measured in weeks). Based on the available covariates, we adopt the following model -
\begin{align}\label{schiz1}
Y_{ij} & = k \;\; \text{if} \;\; \gamma_{k-1} \leq Z_{ij} < \gamma_k, \;\; k = 1,\dots,4, \nonumber \\ Z_{ij} & = \text{treat}_i\beta_1 + t_{ij}\beta_2 + \epsilon_{ij},
\end{align}
where $\epsilon_{ij} (i.i.d) \sim N(0,1)$, and $t_{ij} = \sqrt{time_{ij}}$. Similarly we consider to copulas to model the dependency across time of the ordinal response variables. Here we consider $AR(1)$ structure of the correlation matrix $\mathbf{\Sigma}$ and equal value of $\mathbf{\mu}$ across time points. 
%% The final table. %%
\begin{table}
    \centering
    \begin{small}
    \scalebox{1.0}{
    \tabcolsep = 0.18cm
    \begin{tabular}{|c|c c c c c c|}
    \hline
    \multicolumn{1}{|c|}{Marginal} & \multicolumn{6}{c|}{} \\
    \hline
    \textbf{Parameters} & $\beta_1$ & $\beta_2$ & $\gamma_1$ & $\gamma_2$ & $\gamma_3$ & \\
    \hline
    Est. & -0.4082 & -0.6584 & -2.6283 & -1.4228 & -0.5880 & \\
    SE & 0.1207 & 0.0354 & 0.1243 & 0.1203 & 0.1196 & \\
    \hline
    \multicolumn{1}{|c|}{Copula} & \multicolumn{3}{c|}{GSN} & \multicolumn{3}{c|}{Gaussian} \\
    \hline
    \multicolumn{1}{|c|}{\textbf{Parameters}} & \multicolumn{1}{c}{$p$} & \multicolumn{1}{c}{$\xi$} & \multicolumn{1}{c|}{$\bar{\mu}$} & \multicolumn{3}{c|}{$\xi$} \\
    \hline
    \multicolumn{1}{|c|}{Est.} & \multicolumn{1}{c}{0.8616} & \multicolumn{1}{c}{0.6843} & \multicolumn{1}{c|}{1.4479} & \multicolumn{3}{c|}{0.5423} \\
    \multicolumn{1}{|c|}{SE} & \multicolumn{1}{c}{0.0957} & \multicolumn{1}{c}{0.0959} & \multicolumn{1}{c|}{0.5369} & \multicolumn{3}{c|}{0.0339} \\
    \hline
    \multicolumn{1}{|c|}{Comp-like} & \multicolumn{3}{c|}{\textbf{-4156.80}} & \multicolumn{3}{c|}{-4161.32} \\
    \multicolumn{1}{|c|}{CLAIC} & \multicolumn{3}{c|}{\textbf{8339.26}} & \multicolumn{3}{c|}{8343.17} \\
    \hline
    \multicolumn{7}{|c|}{$D_{12}$ = 0.0147, 95\% = (-0.0081,0.0375), p-val = \textbf{0.2075}} \\
    \hline
    \end{tabular}}
    \end{small}
    \caption{Fitting of schizophrenia data under model (\ref{schiz1}) with GSN and Gaussian copula. Observed composite log-likelihoods, CLAICs and the summary of Voung's statistic are reported.}
    \label{tab:realord1}
\end{table}
\par Results are presented in Table \ref{tab:realord1} for model (\ref{schiz1}) with the GSN and the Gaussian copula. The observed composite log-likelihoods, CLAICs and the summary of Voung's statistic for comparison of two models, are also displayed. Though the CLAIC of the GSN copula based model is less than that of the Gaussian copula based model but the confidence interval of $D_{12}$ contains zero. Hence the two models are not significantly different for this data set. Both of the covariates of this model are significant. The patients under the treatment group in general had improvements over the study period from response $4$ to response $1$, which is shown by the negative value of $\beta_1$. The copula underneath this ordinal longitudinal data is positively skewed as shown by the estimate of $\bar{\mu}$.
%% Declare the end of this section. %%
\section{Discussion}\label{sec9}
In recent times, modeling dependence across time in repeated measurement data has emerged as a crucial area of research. In this article, we introduce a new asymmetric multivariate copula, derived from the multivariate geometric skew-normal distribution proposed by \citet{kundu2017multivariate}. Unlike the skew-normal copula based on Azzalini's skew-normal distribution, the parametric structure of the geometric skew-normal copula is simpler and closed under marginalization. The multivariate Gaussian copula can be viewed as a special case of the geometric skew-normal copula. We examine the dependence properties of the proposed geometric skew-normal copula and provide explicit forms for standard dependence measures such as Kendall's tau and Spearman's rho. Estimating the parameters of the unrestricted geometric skew-normal copula poses challenges, especially for moderate to high dimensional data, due to numerical instabilities during likelihood optimization. To address this, we propose using a block-coordinate ascent algorithm to compute the maximum likelihood estimators (MLEs) of the unknown parameters of both the multivariate geometric skew-normal distribution and the geometric skew-normal copula. Our observations indicate that the proposed algorithm performs efficiently in both cases.
\par The second major contribution of this article is the development of regression models tailored for both continuous and discrete longitudinal data. Leveraging the marginalization property of the GSN copula, we devised a composite likelihood approach to estimate parameters within an ordered probit model, where temporal dependence is characterized by the GSN copula. Our methodologies were rigorously examined through simulation studies and validated against two real-world datasets. Our results indicated that the GSN copula outperforms the Gaussian copula in terms of fit. The geometric skew-normal distribution exhibits significant potential across various statistical modeling applications. However, we noted that the GSN copula lacks non-zero tail dependence, a trait it shares with Azzalini's skew-normal copula. Therefore, there's a pressing need to develop a skew-$t$ extension for both the distribution and the copula, particularly for applications in finance and risk management. Additionally, exploring Bayesian procedures for parameter estimation of the GSN copula stands as a promising avenue for our future research endeavors in multivariate data dependence modeling. \\
%\newpage
%% The Paper ends here. %%
\\
\textbf{Acknowledgment}: The author expresses his sincere gratitude to the Editor in Chief of Journal of Statistical Theory and Practice and two anonymous referees whose constructive comments led to an improved version of the manuscript.
\\ \\
\textbf{Availability of programs}: The R code used to conduct this study is available upon request from the author.
\\ \\
\textbf{Conflict of interest}: The author declares no conflict of interest. \\
%% \textbf{Availability of codes}: The R programs used in this study are available upon request from the author. \\ \\
%% \textbf{Data availability statement}: All data sets used in this study are 
%% The bibliography. %%
\bibliographystyle{agsm}
\bibliography{ref3}
%% %% %% %%
\appendix
\section{Appendix}\label{app3}
\par \textbf{Proofs for propositions \ref{kend1} and \ref{sper1}:} From the definition in \ref{rep1}, we have $\mathbf{X} \overset{d}{=} \sum_{i=1}^N \mathbf{Y_i}$ and $\mathbf{X}' \overset{d}{=} \sum_{i=1}^{N'} \mathbf{Y_i}'$ where $N,N' \sim GE(p)$ and $\mathbf{Y_i},\mathbf{Y_i}' \sim N_2(\mathbf{\mu,\Sigma})$ are all independent. From the expression in (\ref{kendpr1}) we obtain Kendall's tao by
\begin{equation}
P(\mathbf{X} - \mathbf{X}' < 0) = \sum_{n=1}^\infty\sum_{n'=1}^\infty P(\mathbf{X} - \mathbf{X}' < 0|N = n, N' = n')\cdot P(N = n)\cdot P(N' = n'). \nonumber
\end{equation}
Now, $\mathbf{Z} := \mathbf{X} - \mathbf{X}'|N = n, N' = n' \sim N_2\big((n-n')\mathbf{\mu},(n+n')\mathbf{\Sigma}\big)$ and hence
\begin{align}
P(\mathbf{Z} < 0|N = n,N' = n') & = \int_{-\infty}^0\int_{-\infty}^0 f(z_2|z_1)f(z_1)dz_1dz_2 \nonumber \\ & = \int_{-\infty}^0 \Phi\Big(\frac{(n'-n)\mu_2 - \rho(z_1 - (n-n')\mu_1)}{\sqrt{(n+n')(1-\rho^2)}}\Big)\phi\Big(\frac{z_1 - (n-n')\mu_1}{\sqrt{n+n'}}\Big)dz_1 \nonumber \\ & = \int_{-\infty}^{\frac{(n'-n)\mu_1}{\sqrt{n+n'}}} \Phi\Big(\frac{(n'-n)\mu_2}{\sqrt{(n+n')(1-\rho^2)}} - \frac{\rho y}{\sqrt{1 - \rho^2}}\Big)\phi(y)dy, \nonumber
\end{align}
since $Z_2|Z_1 = z_1,N = n,N' = n' \sim N\big((n-n')\mu_2 + \rho(z_1 - (n-n')\mu_1),(n+n')(1 - \rho^2)\big)$.
Similarly we obtain Spearman’s rho by
\begin{equation}
P(\mathbf{X}^* - \mathbf{X} < 0) = \sum_{n^*=1}^\infty\sum_{n=1}^\infty P(\mathbf{X}^* - \mathbf{X} < 0|N^* = n^*, N = n)\cdot P(N^* = n^*)\cdot P(N = n). \nonumber
\end{equation}
Here, $\mathbf{Z}^* := \mathbf{X}^* - \mathbf{X}|N^* = n^*, N = n \sim N_2\big((n^*-n)\mathbf{\mu},n^*\mathbf{I} + n\mathbf{\Sigma}\big)$ and hence
\begin{align}
P(\mathbf{Z}^* < 0|N^* = n^*,N = n) & = \int_{-\infty}^0\int_{-\infty}^0 f(z_2|z_1)f(z_1)dz_1dz_2 \nonumber \\ & = \int_{-\infty}^0 \Phi\Big(\frac{(n-n^*)\mu_2\sqrt{n^*+n} - \frac{n\rho(z_1 - (n^*-n)\mu_1)}{\sqrt{n^*+n}}}{\sqrt{(n^*+n)^2 - n^2\rho^2}}\Big)f(z_1)dz_1 \nonumber \\ & = \int_{-\infty}^{\frac{(n-n^*)\mu_1}{\sqrt{n^*+n}}} \Phi\Big(\frac{(n-n^*)\mu_2\sqrt{n^*+n}}{\sqrt{(n^*+n)^2 - n^2\rho^2}} - \frac{n\rho y}{\sqrt{(n^*+n)^2 - n^2\rho^2}}\Big)\phi(y)dy, \nonumber
\end{align}
since $Z^*_2|Z^*_1 = z_1,N^* = n^*,N = n \sim N\big((n^*-n)\mu_2 + \frac{n\rho(z_1 - (n^*-n)\mu_1)}{n+n^*},n+n^* - \frac{n^2\rho^2}{n+n^*}\big)$. \\ \\
%% %% %% %%
\textbf{Proof of theorem \ref{theorem1}:} Let $\mathbf{U} = (U_1,\dots,U_d)^\intercal \sim C_{d,GSN}(p,\mathbf{\mu,\Sigma})$ and take $x_j = F(u_j|\mu_j,i,p)$, for $j = 1,\dots,d$. Then
\begin{equation}
\mathbf{X} = (X_1,\dots,X_d)^\intercal \sim MGSN_d(p,\mathbf{\mu,\Sigma}). \nonumber
\end{equation}
Therefore, for any permutation $r \in \Gamma$ we have
\begin{equation}
\mathbf{X}_r = (X_{r(1)},\dots,X_{r(d)})^\intercal \sim MGSN_d(p,\mathbf{\mu}_r,\mathbf{\Sigma}_r) \nonumber
\end{equation}
where $\mathbf{\mu}_r$ and $\mathbf{\Sigma}_r$ are the corresponding rearrangements of $\mathbf{\mu}$ and $\mathbf{\Sigma}$ respectively. The moment generating function of $\mathbf{X}_r$ is given as
\begin{equation}
M_{\mathbf{X}_r}(\mathbf{t}) = \frac{p\exp{\big[\mathbf{\mu}_r^\intercal\mathbf{t} + \frac{1}{2}\mathbf{t}^\intercal\mathbf{\Sigma}_r\mathbf{t}\big]}}{1 - (1-p)\exp{\big[\mathbf{\mu}_r^\intercal\mathbf{t} + \frac{1}{2}\mathbf{t}^\intercal\mathbf{\Sigma}_r\mathbf{t}\big]}}. \nonumber
\end{equation}
Since $\mathbf{\Sigma}$ is an exchangeable correlation matrix (i.e. all the off diagonal entries are identical) the quadratic form $\mathbf{t}^\intercal\mathbf{\Sigma}_r\mathbf{t} = \mathbf{t}^\intercal\mathbf{\Sigma}\mathbf{t}$ for all $\mathbf{t}$ and hence $M_{\mathbf{X}_r}(\mathbf{t}) = M_{\mathbf{X}}(\mathbf{t})$, if and only if $\mu_j = \mu \in \mathcal{R}$ for all $j = 1,\dots,d$. Now the univariate geometric skew-normal density follows
\begin{equation}
f_X(x|\mu,1,p) = f_X(-x|-\mu,1,p), \quad \mu \in \mathcal{R}, 0 < p \leq 1. \nonumber
\end{equation}
Hence, $X_i \overset{d}{=} -X_i$ if and only if $\mu = 0$ and that completes the proof. \\ \\
%% %% %% %%
\textbf{Proof of theorem \ref{theorem2}:} Note that for MGSN distribution in (\ref{cdf1}),
\begin{equation}
f_{\mathbf{X}}(\mathbf{x}|\mathbf{\mu,\Sigma},p) = f_{\mathbf{X}}(-\mathbf{x}|\mathbf{-\mu,\Sigma},p) \nonumber
\end{equation}
which is also true for Azzalini's skew-normal and skew-t distribution. Therefore coefficient of upper tail dependence is determined by the lower one. For the bivariate case assume $\mu_1 \geq \mu_2$, which implies $F_1(x) \geq F_2(x)$ for small values of $x$. Hence the lower tail dependence coefficient
\begin{align}
\lambda_L(C_{GSN}) & = \lim_{u \to 0+} \frac{P(F_1(X_1) \leq u,F_2(X_2) \leq u)}{P(F_2(X_2) \leq u)} \nonumber \\ & = \lim_{x \to -\infty} \frac{P(F_1(X_1) \leq F_2(x),F_2(X_2) \leq F_2(x))}{P(F_2(X_2) \leq F_2(x))} \nonumber \\ & \leq \lim_{x \to -\infty} \frac{P(X_1 \leq x,X_2 \leq x)}{P(X_2 \leq x)} \nonumber \\ & = \lim_{x \to -\infty} P(X_1 \leq x|X_2 \leq x) \nonumber \\ & = \lim_{x \to -\infty} \sum_{n=1}^\infty P(X_1 \leq x|X_2 \leq x,N = n)\cdot P(N = n) \nonumber \\ & = \sum_{n=1}^\infty p(1-p)^{n-1} \lim_{x \to -\infty} \frac{\Phi_2(x\mathbf{1}_2|n\mathbf{\mu},n\mathbf{\Sigma})}{\Phi(x|n\mu_2,n)}. \nonumber
\end{align}
Now by applying L’Hospital’s rule
\begin{align}
\lim_{x \to -\infty} \frac{\Phi_2(x\mathbf{1}_2|n\mathbf{\mu},n\mathbf{\Sigma})}{\Phi(x|n\mu_2,n)} & = \lim_{x \to -\infty} \frac{\int_{-\infty}^x \phi_2((x_1,x)^\intercal|n\mathbf{\mu},n\mathbf{\Sigma})dx_1}{\phi(x|n\mu_2,n)} \nonumber \\ & = \lim_{x \to -\infty} \int_{-\infty}^x \frac{1}{\sqrt{2\pi n(1 - \rho^2)}}\exp{[-\frac{(x_1 - n\mu_1 - \rho(x - n\mu_2))^2}{2n (1-\rho^2)}]}dx_1 \nonumber \\ & = \lim_{x \to -\infty} \Phi\Big(\frac{x(1-\rho) - n(\mu_1 - \rho\mu_2)}{\sqrt{n (1-\rho^2)}}\Big) = 0. \nonumber
\end{align}
The reversed inequality can be treated in a similar way. Consequently we get $\lambda_U(C_{GSN}) = 0$ and that completes the proof.
\end{document}